\documentclass[aps,pra,10pt,twocolumn,superscriptaddress]{revtex4-1}
\usepackage{graphicx,hyperref,doi}

\renewcommand{\vec}{\mathbf}

\begin{document}

\title{Comparing many-body approaches against the helium atom exact solution}

\author{Jing Li}

\affiliation{Universit{\'e} Grenoble Alpes, 38000 Grenoble, France}

\affiliation{CNRS, Institut N\'eel, 38042 Grenoble, France}

\author{N.\ D.\ Drummond}

\affiliation{Department of Physics, Lancaster University, Lancaster
  LA1 4YB, United Kingdom}

\author{Peter Schuck}

\affiliation{Universit{\'e} Grenoble Alpes, 38000 Grenoble, France}

\affiliation{CNRS, LPMMC, 38042 Grenoble, France}

\affiliation{CNRS, Institut de Physique Nucl\'eaire, IN2P3, Universit\'e Paris-Sud, 91406 Orsay, France}

\author{Valerio Olevano}

\email{valerio.olevano@grenoble.cnrs.fr}

\affiliation{Universit{\'e} Grenoble Alpes, 38000 Grenoble, France}

\affiliation{CNRS, Institut N\'eel, 38042 Grenoble, France}

\affiliation{European Theoretical Spectroscopy Facility (ETSF)}


\date{\today}

\begin{abstract}
Over time, many
different theories and approaches have been developed to tackle the
many-body problem in quantum chemistry, condensed-matter physics, and
nuclear physics.  Here we use the helium atom, a real system rather than a model,
and we use the exact solution of its Schr\"odinger equation as a
benchmark for comparison between methods.  We present new results 
beyond the random-phase approximation (RPA) from a renormalized RPA
(r-RPA) in the framework of the self-consistent RPA (SCRPA)
originally developed in nuclear physics, and compare them with
various other approaches like configuration interaction
(CI), quantum Monte Carlo (QMC), time-dependent density-functional
theory (TDDFT), and the Bethe-Salpeter equation on top of the $GW$
approximation.  Most of the calculations are consistently done on the
same footing, e.g.\ using the same basis set, in an effort for a most
faithful comparison between methods.
\end{abstract}

\maketitle

\section{Introduction}

The neutral helium atom and other two-electron ionized atoms are among
the simplest many-body systems in nature.  Here ``many-body'' is reduced
to only three bodies, two electrons plus the nucleus. 
Even when treating the nucleus classically 
(i.e., as an external classical source and neglecting its wave function), 
in quantum mechanics two interacting bodies (the two electrons) already
raise a \textit{many-body} problem (an inhomogeneous two-body problem in helium).
The Schr\"odinger equation cannot be solved in a closed form. 
The calculation of many-body correlation energies and correlation effects
presents similar difficulties in two-electron systems 
(including the noteworthy case of the hydrogen molecule)
as in any other many-body system.

Nevertheless, thanks to the pioneering  work of Hylleraas in 1929 \cite{Hylleraas29}, the helium atom (and two-electron atoms in general) is 
almost a unique case in which we own an \textit{exact} solution,
though not in a closed form.
By exploiting the full rotational symmetry of the system and rewriting
the Schr\"odinger equation in reduced degrees of freedom,
these being the three scalar Hylleraas coordinates over which the wave function
is expanded as a power series, a numerical solution can be found.
This numerical solution is ``exact'' in the sense that it consists of a number and a quantifiable margin of error on that number, together
with the possibility of arbitrarily reducing that margin of error.
The historical series of published results
\cite{Kinoshita57,FrankowskiPekeris66,ThakkarKoga94,Goldman98,
Drake99,SimsHagstrom02,DrakeNistor02,Korobov02,Schwartz06,NakashimaNakatsuji07}
(see Table I in \cite{NakashimaNakatsuji07}) has confirmed the numerically exact nature of the Hylleraas method for helium.

Hylleraas's original solution had a relative error of only $10^{-4}$,
which is remarkable for a time in which computers did not yet exist.
It played a fundamental role in assessing the validity of quantum
mechanics as a universal theory that does not just apply to the
hydrogen atom. Once higher-order effects are taken into account, such
as nuclear finite-mass recoil (reduced mass of the electron and mass
polarization term), relativistic fine structure (e.g.\ relativistic
correction to the velocity, spin-orbit coupling, etc.), and quantum
electrodynamic (QED) radiative corrections (the analog of the Lamb
shift of hydrogen)
\cite{PachuckiSapirstein03,PachuckiYerokhin10,ZhangDrake96}, its
quantitative agreement with the experiment, within the measured and
calculated error bars, was one of the first triumphs of
quantum mechanics \cite{FrankowskiPekeris66}.

Over the years, the Hylleraas calculation has been improved more
and more \cite{Kinoshita57,FrankowskiPekeris66,ThakkarKoga94,Goldman98,
Drake99,SimsHagstrom02,DrakeNistor02,Korobov02,Schwartz06,NakashimaNakatsuji07},
reaching an accuracy of 35 decimal digits in 2006
\cite{Schwartz06}, a result confirmed and further extended \cite{NakashimaNakatsuji07}, which required computer octuple
precision.  Beyond the academic interest, the comparison of such an
accurate theoretical result with experimental measurements of the helium
excitation spectrum has been proposed to estimate the fine
structure constant accurately.

The availability of an exact solution suggests that the helium
atom can serve as a workbench for many-body theories. 
Many body theories were at the beginning mostly tailored for systems with many, 
up to infinite particles. More recently one requires that a good 
many body approach embraces simultaneously the small and high number of particles cases. 
A two-electron atom might appear a limiting case to study the many-body problem.
However, it is not that far from the hydrogen molecule, of interest in
molecular physics and quantum chemistry, or the deuterium nucleus, of interest in nuclear physics.
Each of these systems presents a 
nontrivial many-body problem to describe the electronic correlation beyond
the Hartree-Fock (HF) exchange. 
Many different formalisms beyond HF
have been developed over time aimed at the solution of the
many-body problem.  While exact in principle, in
practice all approaches rely on approximations and recipes whose
validity are difficult to judge.  The general tendency is to evaluate
them against experiment.  However, benchmarks against experiment are
always affected by unaccounted effects not present in the theoretical
description (non-Born-Oppenheimer, electron-phonon, relativistic
corrections, etc.), which can mask the real many-body performances of
the approaches.  Validation of many-body approaches against the
benchmark of an exact solution is an unavoidable step for further
improvement.  When calling for an exact solution one first thinks of a
model system.  Workbenches for many-body theories have been identified
in more or less realistic simplified models, e.g.\ by replacing the
long-range $1/r$ real electromagnetic interaction by a local
interaction $\delta(r)$, and/or by discretizing the space, or somehow
reducing the number of degrees of freedom of the system.  
One example of particular relevance here is the \textit{spherium} model.
With respect to the helium atom, in spherium only the angular degrees of freedom are considered, whereas the radial ones are dropped by confining the two electrons on the 2D surface of a sphere of radius $R$.
However, the interaction is the real 3D $1/r$ across the sphere.
$R$ is a model parameter which allows the tuning of the electronic density (like $r_s$ of the jellium model) and so of the correlations: this allows interesting studies which are impossible in real helium.
By comparing the spherium solution to the real helium atom electronic structure (Fig.~\ref{heelecstruct}) one can appreciate the validity of this model to describe nature, as well as its limits.
An interesting work on spherium also comparing many-body approaches is Ref.~\cite{LoosBerger18}.
However the exact solution is often unknown even for simple models, or known only
in particular cases or in reduced dimensions.  Another drawback is
that many-body theories could be checked on unrealistic features of
models, and so one theory can be validated with respect to another on
aspects that might be absent in real systems.  So, we think that the
helium atom and its exact solution is certainly preferable to more schematic 
models
as a benchmark for many-body theories.  Furthermore, the electronic
structure of helium is very rich (see Fig.\ \ref{heelecstruct}),
presenting a complex spectrum of many excitations of different nature; it is certainly much more critical
for a theory to be able to reproduce, as a whole, such a rich
electronic structure rather than a model that can present just a
couple of levels.
Finally, the helium atom represents a very severe workbench
for testing condensed-matter approaches devised for describing correlations in 
infinite solids by e.g.\ the introduction of the concept of ``screening,'' 
a check that, according to our results, these approaches have surprisingly passed.

\begin{figure}
 \includegraphics[width=\columnwidth]{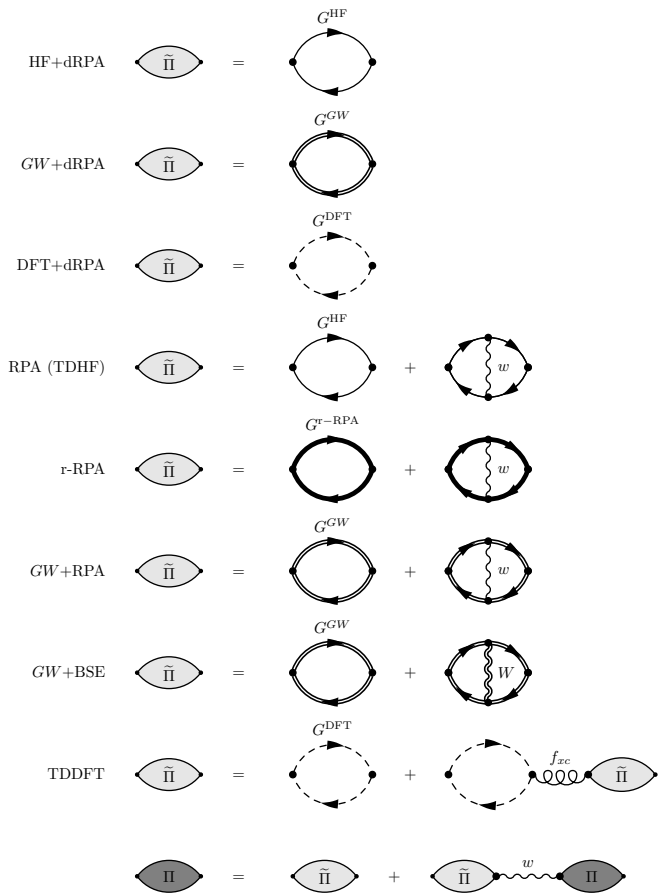}
 \caption{Irreducible polarizability $\widetilde{\Pi}$ in the various approximations studied in this work.
          HF+dRPA: direct RPA (dRPA) or ring approximation on top of the Hartree-Fock (HF) electronic structure;
          $GW$+dRPA: dRPA on top of the $GW$ electronic structure;
          DFT+dRPA: dRPA on top of the (either exact or approximated, e.g., LDA, GGA, etc.) DFT Kohn-Sham electronic structure;
          RPA (TDHF): random-phase approximation approximation, also known as linearized time-dependent Hartree-Fock;
          r-RPA: renormalized RPA;
          $GW$+RPA: RPA on top of $GW$;
          $GW$+BSE: Bethe-Salpeter equation on top of $GW$;
          TDDFT: linear response time-dependent density-functional theory with kernel $f_{xc}$ on top of either exact or approximated (e.g.\ LDA) DFT.
          Last line: Dyson equation $\Pi = \widetilde{\Pi} + \widetilde{\Pi} w \Pi$ between the irreducible $\widetilde{\Pi}$ and the reducible polarizability $\Pi$.
          The wiggly line marked $w$ indicates the bare many-body interaction (here the Coulomb interaction), while the double wiggly line indicates the screened interaction $W$.
 }
 \label{polarizabilities}
\end{figure}

\begin{figure*}
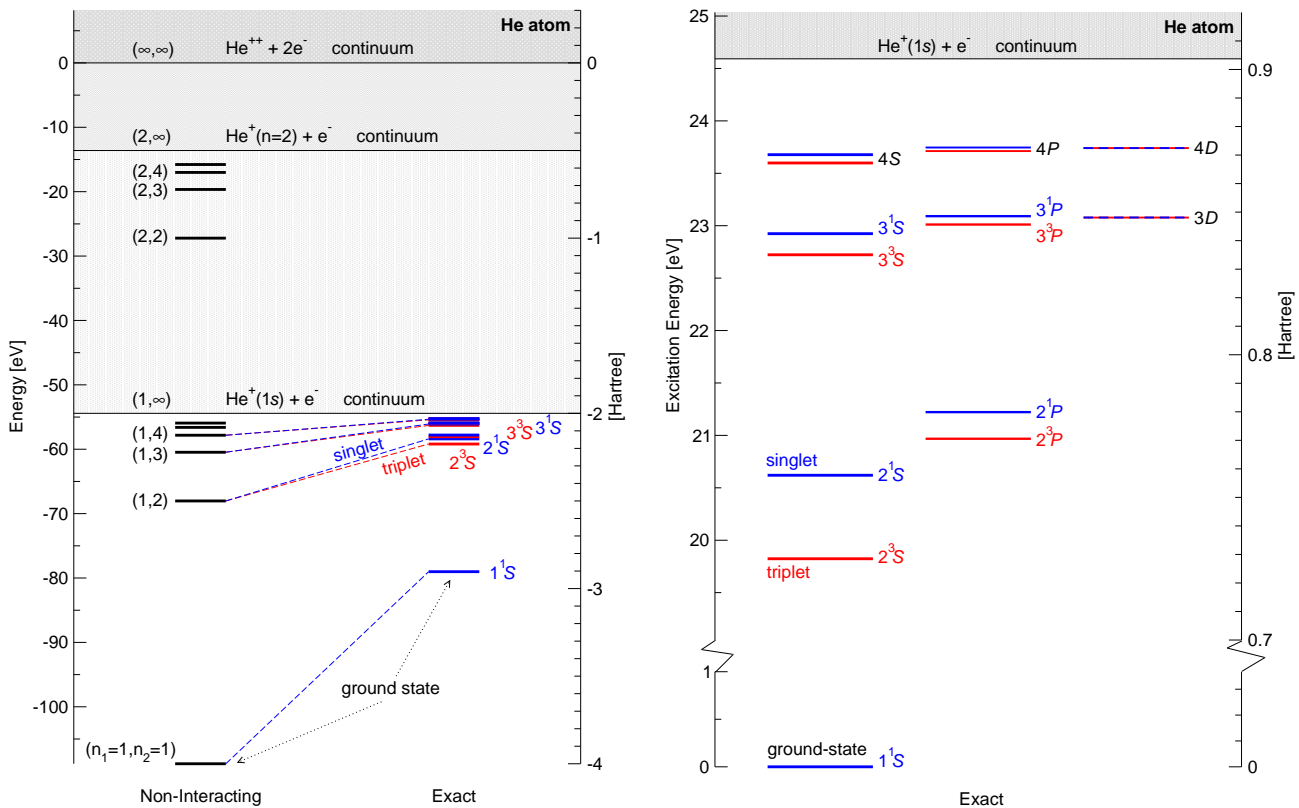

 \includegraphics[height=0.45\textheight]{heindpartexact}
  \hfil
 \includegraphics[height=0.45\textheight]{heexact}
 \caption{Helium atom full electronic structure (left panel). Both the
   noninteracting, independent-particle spectrum [Eq.~(\ref{IP})]
   and the exact \cite{KonoHattori84} spectrum are shown.
   The right panel is a zoom on the first
   excitations of the exact spectrum.}
 \label{heelecstruct}
\end{figure*}

In the present work, the intention is to perform a comparison of several 
many body approaches. Most of those approaches are well known in
condensed-matter physics. However, a direct comparison of their performances
is often hampered by not consistent techniques of numerical resolution. 
One objective of the present work, therefore, is to improve on this.
Second, we also want to introduce and apply a method used in nuclear physics
which is the equation of motion (EOM) approach to go beyond the standard 
random-phase approximation (RPA)\@. 
It is called the self-consistent RPA (SCRPA), of which the 
renormalized RPA (r-RPA) is a sub-product \cite{CataraVanGiai96,CataraSambataro98}.
We will give a short outline of this approach. 
We right now clarify that all along this paper we consistently use the 
nuclear physics convention to define the \textit{random-phase approximation}
(RPA) which in quantum chemistry and condensed-matter physics
is rather known as linearized time-dependent Hartree-Fock (TDHF)\@.
The RPA here contains both the direct and the exchange terms, and 
should not be confused with the RPA of condensed-matter physics
(also known as the \textit{ring} approximation),
which only contains the direct term.
We will here refer to the latter as dRPA (direct RPA) to avoid confusion.
In order to provide the reader with an orientation table 
among the acronyms used in this article,
in Fig.~\ref{polarizabilities} we present the Feynman diagrams
for the irreducible polarizability $\widetilde{\Pi}$ corresponding
to all the approximations explored in this work.
The last line of Fig.~\ref{polarizabilities} presents the 
Dyson equation relating the irreducible $\widetilde{\Pi}$ to
the reducible polarizability $\Pi$ whose poles are the excitation
energies tabulated in this work for helium.

In order to compare with the other approaches, we at the same time
calculate helium ground and excited states by some of the most
widespread many-body approaches, including Hartree-Fock (HF), quantum
Monte Carlo (QMC), quantum chemistry configuration interaction (CI),
density-functional theory (DFT) and time-dependent density-functional
theory (TDDFT), Bethe-Salpeter equation (BSE)
\cite{SalpeterBethe51,HankeSham79,OnidaAndreoni95} on top of the $GW$
approximation
\cite{Hedin65,StrinatiHanke80,StrinatiHanke82,HybertsenLouie85,GodbySham87},
and the dRPA approximation 
on top again of the $GW$ electronic structure, or also of the HF or the DFT ones
(see Fig.~\ref{polarizabilities}).
Some of these results were previously
presented in the literature, but here we made the effort to recalculate
most of them on the same footing, in particular using the same
Gaussian basis set, which, as we will see, significantly affects the
accuracy of the results. This yields a more faithful comparison between
methods.  
\footnote{One should expect that, on localized gaussians, long-range $1/r$ methods (e.g., wavefunction based, GW, hybrids) converge more slowly than short-range $e^{-r}$ methods (e.g., LDA xc-potential or other DFT pure functionals) \cite{WangWilson04}: this is the reason why we chose basis-set families optimized for correlation wavefunction methods. 
Another possible way is to assess each method with its best complete basis set.
However, if one checks Fig. 1 of Ref.~\cite{Jensen17}, one can see that for He the convergence error can be reduced from $10^{-4}$ to $10^{-6}$ Ha using basis-set families optimized for DFT LDA.
Since the error of the LDA approximation with respect to the exact Hylleraas is already $7 \cdot 10^{-2}$ Ha (Table~\ref{groundstate}), the basis-set convergence error of $10^{-4}$ can be neglected.
}
For obvious reasons, only the QMC calculations and the
exact-DFT (including also TDDFT on top of exact-DFT) calculations,
apart of course from the exact Hylleraas calculation, are not based on
the Gaussian basis set.  Most importantly, the spirit which has
driven our comparison of so many methods was to understand and
demonstrate the \textit{effective} performances achieved \textit{in
  practice} by a given methodology, avoiding idealistic 
statements.  One can
claim, for example, that ``solving full Hedin equations
self-consistently will provide the exact solution,'' but this remains
an abstract statement if nobody was ever able to perform such a
calculation for any real or even model system.  When going for a real
calculation one can face unforeseen problems related to, for
example, basis-set issues, nonlinearity of equations, divergences to
be avoided, self-consistency instabilities, etc., which can reduce the
exactness of the solution achieved, if not actually preventing the
achievement of a solution.  We will discuss such issues in the present
work.  Beyond this, to situate the performance and pros and cons of each
approach with respect to the others, the purpose of this article is to
propose a workbench and a methodology for evaluation of future
developments.

Most of the results shown in this paper are new and to our knowledge have not
been presented earlier in the literature:
i) r-RPA calculations have
so far only been applied to models in use in nuclear physics and 
to the jellium-sphere model \cite{CataraVanGiai96,CataraSambataro98},
but not previously to real systems. 
We also give a short outline of this method, as it is not well known outside the
nuclear physics community.
ii) The novelty aspect is also apparent for our d-RPA calculations, which
are applied on top of three approximations, namely HF, $GW$, and DFT-LDA\@.
iii) Likewise for the $GW$, $GW$+RPA, and $GW$+BSE results studied as a function
of starting point, from PBE \cite{PerdewErnzerhof96} to full 
HF exchange (PBEh) \cite{AtallaScheffler13}.
iv) Although the DFT-LDA + TDLDA helium-atom excitation spectrum
has been discussed several times in the literature \cite{PetersilkaBurke00,WassermanBurke03,StonerGorling01,Maitra16}, numerical results 
have never been published to our knowledge. We fill this gap in this work.
v) Finally, our variational (VMC) and diffusion Monte Carlo (DMC) calculations present improved results
with respect to earlier QMC works on helium, 
and the achievement of an accuracy high enough to be at the level
of the experimental error bar.

The paper is organized as follows: we first introduce the electronic
structure of helium and the exact Hylleraas solution, showing this to
be a safe reference.  We then describe in particular
the SCRPA approach, referring to the literature for the other well known methods, and
present the parameters of all our calculations.  The results will be
presented, first for the ground state and then for the excited states.
Our conclusions are drawn at the end.  We will generally use
atomic units (Hartree, Bohr), but will also report energies in
electronvolts (eV) when this is more intuitive.  The zero of the
energies will be fixed at the helium atom double excitation level
He$^{++} + 2 e^-$ when studying the helium ground state, and at the
ground state $1^1\!S$ when studying the excitation spectrum.

\section{Helium atom electronic structure and exact solution}

The experimental spectrum of a real helium atom is affected by many
effects (e.g.\ the finite mass of the nucleus, relativistic corrections,
and QED radiative corrections) beyond many-body correlations.  These
effects are small corrections \cite{FrankowskiPekeris66} that can be
calculated at the first order, but must be taken into account in a
comparison with experiment within experimental and theoretical
error bars.  Here we are interested in reproducing not the experiment,
but an exact solution as a benchmark for many-body theories and their
performances on correlation.  So our workbench system will be an
idealized nonrelativistic helium atom, with infinite nuclear mass and
without relativistic and QED effects, whose Hamiltonian is
\begin{equation}
 H = -\frac{\partial^2_{r_1}}{2} -\frac{\partial^2_{r_2}}{2}
 -\frac{Z}{r_1} - \frac{Z}{r_2} + \frac{1}{|\mathbf{r}_1 - \mathbf{r}_2|} , \label{H}
\end{equation}
consisting of the kinetic terms, the interaction with the nucleus of charge
$Z$, and the two-body Coulomb interaction between the electrons (last
term).  If we neglect the latter (noninteracting or
independent-particle approximation) the Hamiltonian can be split into
two single-particle Hamiltonians of hydrogenic form, and the
solution for the excitation spectrum can easily be written
\begin{equation}
 E^0_{n_1 n_2} = - \frac{Z^2}{2} \left( \frac{1}{n_1^2} + \frac{1}{n_2^2} \right)
 . \label{IP}
\end{equation}
This is the noninteracting, independent-particle electronic structure
reported (for helium $Z=2$) in Fig.\ \ref{heelecstruct}, left.  One can
identify the ground state, corresponding to the principal quantum
numbers $(n_1=1,n_2=1)$, the first excitations, $(1,2)$, $(1,3)$,
\dots, forming a Rydberg series up to the first ionization onset
$(1,\infty)$ in which we are left with a He$^+(1s) + e^-$ helium
positive ion in its hydrogenic $1s$ ground state, plus a free
electron.  We then have so-called double excitations $(n_1 > 1,n_2 > 1)$, which
are resonant with the continuum of the first ionization onset, and
further single ionization onsets $(n_1 > 1,\infty)$.  Finally we have the full ionization
level $(\infty,\infty)$, in which we are left with the bare He nucleus
plus two free electrons, He$^{++} + 2 e^-$.

When comparing the independent-particle with the exact electronic
structure (Fig.\ \ref{heelecstruct} left panel), one can see that the
many-body term has an important, non-negligible effect already in
helium.  There are important shifts, especially for the ground state,
and splits of levels according further quantum numbers as the total
spin $S$ and the total orbital angular momentum $L$ (see also
Fig.\ \ref{heelecstruct} right panel).  A good many-body theory should
be able to reproduce reasonably well both shifts and splits.

We now also briefly explain how the exact solution to the
Schr\"odinger equation for the Hamiltonian in Eq.\ (\ref{H}) could be
obtained by Hylleraas.  Starting from the solution of the hydrogen
atom, and exploiting the full rotational symmetry of the ionic
potential (an important simplification with respect to e.g.\ the
hydrogen molecule) it was possible to write the electronic wave
functions $\Psi(s,t,u)$ in terms of only three scalar coordinates,
\begin{eqnarray*}
  s &=& r_1 + r_2 \\
  t &=& r_1 - r_2 \\
  u &=& r_{12} = |\mathbf{r}_1 - \mathbf{r}_2|
\end{eqnarray*}
instead of the two vectors or six scalars $\Psi(\mathbf{r}_1,
\mathbf{r}_2)$ normally required for a two-electron system.  The wave
function is then written as a power series over the $s$, $t$, and $u$
Hylleraas coordinates,
\begin{equation}
 \Psi(s,t,u) = e^{-ks} \sum_{l,m,n} c_{l,m,n} s^l t^m u^n
 , \label{Psi}
\end{equation}
apart from an important cusp factor $e^{-ks}$ in analogy with the
solution of the hydrogen atom.  
It has been demonstrated \cite{Kinoshita57} that the expansion Eq.~(\ref{Psi}), including negative powers $l,m<0$, represents a formal solution to the He Schr\"odinger Eq.~(\ref{H}).
The solution is found variationally,
by minimizing the energy with respect to the free parameters
$c_{l,m,n}$ and $k$.  It is possible to select the symmetry of the
wave function, for example by choosing even $m$ for space-symmetric
singlet solutions and odd $m$ for space-antisymmetric triplets, like
also the orbital character (upon reintroducing angular variables within multiplicative spherical harmonics \cite{KonoHattori84}), and even the principal quantum number.
This provides access not only to the ground state, but also all
excited states, both energies and wave functions, and so also
oscillator strengths.  The Hylleraas accuracy of $10^{-4}$ (relative error), 
which was obtained with a reduced sum in Eq.\ (\ref{Psi}) running only on
positive powers, was in the following years improved by extending the
series also to include negative powers \cite{Kinoshita57}.  An
important increase in the accuracy was obtained thanks to a better
description of the coalescence region at the origin by introducing a
logarithmic singularity $\ln(s)$ \cite{FrankowskiPekeris66}, like in the
wave functions which allowed Schwartz \cite{Schwartz06} to obtain an
accuracy of 35 decimal digits,
\begin{equation}
 \Psi(s,t,u) = e^{-ks} \sum_{j,l,m,n} c_{j,l,m,n} s^l (t/s)^m (u/s)^n \ln^j(s)
 . \label{Psiln}
\end{equation}
The logarithm factor, first introduced by Frankowski and Pekeris
\cite{FrankowskiPekeris66}, was important to overcome the Kinoshita
\cite{Kinoshita57} accuracy of $10^{-6}$ Ha.  

\section{Formalisms}

In this section we will in particular introduce SCRPA and detail the
r-RPA approach we have followed.  

\subsection{Standard, renormalized and self-consistent RPA}

The standard and also self-consistent RPA equations can be quite
straightforwardly derived from the equations of motion (EOM) \cite{Rowe66,Rowe68}
of excitation creation operators $\hat{Q}^\dag_\lambda$, defined by
\[
 \hat{Q}^\dag_\lambda | \Phi_0 \rangle = | \Phi_\lambda \rangle,
~~{\rm with} ~~~ \hat{Q}^\dag_\lambda =| \Phi_\lambda \rangle \langle \Phi_0 | ,
\]
with $\Phi_\lambda$ the excited states, both singlets and triplets, and  $\Phi_0$ the ground state,
\[
 \hat{H} | \Phi_\lambda \rangle = E_\lambda | \Phi_\lambda \rangle
 ,
\]
of the full Hamiltonian $\hat{H}$,
\begin{eqnarray*}
 \hat{H} &=& \hat{H}^0 + \hat{W} = \\ &=& \sum_{k_1k_2}
 \epsilon^0_{k_1 k_2} \hat{c}^\dag_{k_1} \hat{c}_{k_2} + \frac{1}{4}
 \sum_{k_1 k_2 k_3 k_4} \bar{v}_{k_1 k_2 k_3 k_4} \hat{c}^\dag_{k_1}
 \hat{c}^\dag_{k_2} \hat{c}_{k_4} \hat{c}_{k_3} ,
\end{eqnarray*}
where $\hat{H}^0 = \hat{T} + \hat{V}_\mathrm{ext}$ is the
noninteracting Hamiltonian and $\epsilon^0_{kk'}$ is its matrix elements
with respect to a basis set $\phi_k(r)$ over which we also define the
creation/annihilation operators $\hat{c}^\dag_k$/$\hat{c}_k$, while
\begin{equation}
 \bar{v}_{k_1 k_2 k_3 k_4} = \langle \phi_{k1} \phi_{k2} | v |
 \phi_{k3} \phi_{k4} \rangle - \langle \phi_{k1} \phi_{k2} | v |
 \phi_{k4} \phi_{k3} \rangle
 \label{RPAkernel}
\end{equation}
are the antisymmetrized matrix elements of the many-body interaction
$v$, in this work the Coulomb interaction $v(\mathbf{r},\mathbf{r}') =
1 / |\mathbf{r}-\mathbf{r}'|$.  The Hermitian conjugated annihilation
operators are subject to the killing condition on the ground state,
\begin{equation}
   \hat{Q}_\lambda | \Phi_0 \rangle = 0
 \label{killing} .
\end{equation}
From the equation of motion obeyed by the $\hat{Q}^\dag_\lambda$, we
can derive the equation \cite{RingSchuck,Rowe66,Rowe68}
\begin{equation}
 \langle \Phi_0 | [ \delta \hat{Q} , [ \hat{H}, \hat{Q}^\dag_\lambda ] ] |
 \Phi_0 \rangle = \Omega_\lambda \langle \Phi_0 | [\delta \hat{Q} ,
   \hat{Q}^\dag_\lambda ] | \Phi_0 \rangle
 , \label{EMM}
\end{equation}
where $\Omega_\lambda = E_\lambda - E_0$ are the excitation energies
measured from the ground state, and $\delta \hat{Q}$ is an arbitrary variation of the  operator $\hat{Q}^\dag_\lambda $,
associated to a generic state of the Hilbert space $| \Phi \rangle =
\delta \hat{Q}^\dag | \Phi_0 \rangle$. 
For a variant of the derivation of Eq.~(\ref{EMM}), see Ref.~\cite{JemaiSchuck13}
where the minimization of an energy weighted sum rule is employed.

So far everything is exact.
We understand that $\hat{Q}^\dag_\lambda $ is a complicated many body operator 
which may be considered as  a superposition of one body, two-body, \dots, N-body operators.
We now restrict, as an approximation, the $  \hat{Q}^\dag_\lambda  $ 
operators to be of the one-body form
\[
 \hat{Q}^\dag_\lambda = \sum_{k_1 \ne k_2} \chi^\lambda_{k_1 k_2}
 \hat{c}^\dag_{k_1} \hat{c}_{k_2}
 ,
\]
with both $k_1$ and $k_2$ running over all indices, besides the diagonal configurations. 
We obtain the secular equation
\begin{equation}
\sum_{k'_1k'_2} \mathcal{S}_{k_1k_2,k'_1k'_2}
\chi^{\lambda}_{k'_1k'_2} = \Omega_\lambda \chi^{\lambda}_{k_1k_2}
, \label{SCRPA}
\end{equation}
with the matrix $\mathcal{S}$ defined as
\[
 \mathcal{S}_{k_1k_2,k'_1k'_2} = \langle \Phi_0 | [\hat{c}^\dag_{k_1}
   \hat{c}_{k_2},[\hat{H},\hat{c}^\dag_{k'_1} \hat{c}_{k'_2}]]| \Phi_0
 \rangle (n_{k'_2}-n_{k'_1})^{-1},
\]
where $n_k\delta_{kk'}$ is the single-particle density matrix
\[
 \langle \Phi_0 | \hat{c}^\dag_k \hat{c}_{k'} | \Phi_0 \rangle =
 \delta_{kk'} n_k 
\]
supposed, for convenience, to be diagonal.
(This is an approximation. It can be avoided without formal problems 
by using the canonical basis which diagonalizes the single-particle density matrix,
but usually it does not add much to the accuracy of the solution.)
By developing the double commutator we obtain
\begin{widetext}
\begin{eqnarray*}
  \mathcal{S}_{k_1k_2,k'_1k'_2} &=& ( \epsilon_{k_1} - \epsilon_{k_2}
  ) \delta_{k_1k'_1} \delta_{k_2k'_2} + ( n_{k_2} - n_{k_1} )
  \bar v_{k_1k'_2k_2k'_1} + 
    \\ && 
    \bigg[
    -\delta_{k_2k'_2}\frac{1}{2}\sum_{j_1j_2j_3}\bar
    v_{k_1j_1j_2j_3}C_{j_2j_3k'_1j_1}
    -\delta_{k_1k'_1}\frac{1}{2}\sum_{j_1j_2j_3}\bar
    v_{j_1j_2k_2j_3}C_{k'_2j_3j_1j_2} + \\ && \sum_{j_1j_2}(\bar
    v_{k_1j_1k'_1j_2}C_{k'_2j_2k_2j_1}+\bar
    v_{k'_2j_1k_2j_2}C_{k_1j_2k'_1j_1}) -\frac{1}{2}\sum_{j_1j_2}(\bar
    v_{k_1k'_2j_1j_2}C_{j_1j_2k_2k'_1}+\bar
    v_{j_1j_2k_2k'_1}C_{k_1k'_2j_1j_2})\bigg ](n_{k'_2}-n_{k'_1})^{-1}
  , \label{S}
\end{eqnarray*}
in terms of the cumulant of the two-particle correlation functions
\[
 C_{k_1k_2k_3k_4} = \langle
 \Phi_0|\hat{c}^\dag_{k_3}\hat{c}^\dag_{k_4}\hat{c}_{k_2}\hat{c}_{k_1}|\Phi_0\rangle
 - n_{k_1}n_{k_2}[\delta_{k_1k_3}\delta_{k_2k_4} -
   \delta_{k_2k_3}\delta_{k_1k_4}]
 ,
\]
\end{widetext}
and of the single-particle energies $\epsilon_k$ and basis set
$\phi_k$ eigensolutions of the equation
\begin{equation}
  [ H^0 + V^\mathrm{MF} ] \phi_k = \epsilon_k \phi_k
  , \label{MFeq}
\end{equation}
with the mean-field potential given by 
\begin{equation}
  V^\mathrm{MF}_{k_1k_2} = \sum_k\bar v_{k_1kk_2k}n_k
  . \label{MFpot}
\end{equation}
(Note that \textit{a priori} in the mean-field basis $\phi_k$ neither
the kinetic energy, nor the external potential are diagonal
separately).  The correlation functions $C$ contain only the fully
connected terms of the two-body density matrix, i.e., the fully
correlated part.

The correlation functions $C$ can be expressed by the RPA solution, and
thus Eq.~(\ref{SCRPA}), with the expression for $\mathcal S$ above,
constitute the full self-consistent RPA (SCRPA) equations. 
If we neglect in $\mathcal S$
all two-body correlation functions $C$, we obtain the renormalized RPA (r-RPA) approach. 
Replacing additionally the correlated
$n_k$ by the uncorrelated integer Hartree-Fock occupation numbers,
$n^\mathrm{HF}_h = 1$ for holes and $n^\mathrm{HF}_p = 0$ for
particles, we re-obtain the standard RPA equations with the exchange
term \cite{Rowe66,Rowe68,RingSchuck}
\begin{equation}
  \left(
  \begin{array}{cc}
   A & B \\
   B^* & A^*
  \end{array}
  \right)  
  \left(
  \begin{array}{c}
  X^\lambda \\ Y^\lambda
  \end{array}
  \right) = \Omega_\lambda
  \left(
  \begin{array}{cc}
   1 & 0 \\
   0 & -1
  \end{array}
  \right)  
  \left(
  \begin{array}{c}
  X^\lambda \\ Y^\lambda
  \end{array}
  \right)
  , \label{RPA}  
\end{equation}
with
\begin{eqnarray*}
 A_{ph,p'h'} &=& (\epsilon^\mathrm{HF}_p - \epsilon^\mathrm{HF}_h)
 \delta_{pp'} \delta_{hh'} + (n^\mathrm{HF}_h - n^\mathrm{HF}_p) \bar{v}_{ph'hp'} \\ B_{ph,p'h'} &=&
(n^\mathrm{HF}_h - n^\mathrm{HF}_p) \bar{v}_{pp'hh'} .
\end{eqnarray*}

Indeed in this case the mean-field potential [Eq.\ (\ref{MFpot})] is
exactly the Hartree potential plus the exchange (Fock) operator, and
Eq.\ (\ref{MFeq}) is the Hartree-Fock equation, so that $\epsilon_k$
and $\phi_k(r)$ are the Hartree-Fock energies and wave functions.  So the
$\mathcal{S}$ matrix contains HF energies $\epsilon_k$ along the
diagonal, while the kernel reduces to the $\bar{v}$ terms.

In this work we went beyond standard RPA towards self-consistency, but
did not pursue full SCRPA\@. The latter task remains for the future. 
We followed the r-RPA approach where in Eq.~(\ref{RPA}) 
all HF occupation numbers and energies are replaced by correlated ones, see, e.g.,
 Catara \textit{et al.}\
\cite{CataraVanGiai96,CataraSambataro98}.  In this approach, at each
step of self-consistency a new, beyond Hartree-Fock, correlated
mean-field electronic structure is calculated.  The correlated
electronic structure is characterized by noninteger occupation
numbers $n_h$ and $n_p$, unlike the integer
uncorrelated Hartree-Fock occupation numbers.  The
depletion/repletion with respect to HF uncorrelated occupation numbers
can, e.g., be calculated from the correlated RPA amplitudes $\chi^\lambda_{hp}$
(\textit{number operator} method \cite{Rowe66, CataraVanGiai96})
\begin{eqnarray}
  n_p &=& \frac{1}{2} \sum_{\lambda h} (n_h-n_p) |\chi^\lambda_{hp}|^2
  , \label{occ1} \\
  n_h &=& 1 - \frac{1}{2} \sum_{\lambda p} (n_h-n_p)|\chi^\lambda_{hp}|^2
  . \label{occ2}
\end{eqnarray}
(The same result can be obtained with other formulations \cite{SchuckTohyama16}). 
For small depletion/repletion one can replace the
occupation numbers on the right-hand side with uncorrelated
Hartree-Fock 0/1 occupation numbers.  These expressions are correct to
second order in $|\chi^\lambda_{hp}|$.  Catara \textit{et al.}\
\cite{CataraVanGiai96,CataraSambataro98} considered higher-order
corrections but we will see that in helium the
depletion/repletion of occupation numbers constitute a correction of
less than 1\%, so that higher-order corrections are negligible, and
stopping at second order is safe.  Note that the occupation numbers of
Eqs.\ (\ref{occ1}) and (\ref{occ2}) fulfill Luttinger's theorem, since the
particle number $N$ is conserved:
\[
 \sum_h n_h + \sum_p n_p = N
 .
 \]
Also we will restrict the configuration space to particle-hole (hole-particle).

So, starting from standard RPA, after having solved the RPA equations
and having calculated the $\chi$ amplitudes, we recalculate the correlated
occupation numbers using Eqs.\ (\ref{occ1}) and (\ref{occ2}), the new
mean-field potential using Eq.\ (\ref{MFpot}) and the new correlated
energies $\epsilon_k$ using Eq.\ (\ref{MFeq}).  The procedure is cycled
till self-consistency.  This r-RPA approach can be
considered an approach towards SCRPA with the important simplification
that the two-body correlation functions in the $\mathcal{S}$ matrix are
neglected, but correlations are at least self-consistently introduced
in the occupation numbers and in the single-particle energies that now
depart from the uncorrelated HF expressions (see Fig.~\ref{polarizabilities}).

Finally, this methodology also allows one to calculate the total energy of
the ground state, that is the correlation contribution of RPA,
$E_c^\mathrm{RPA}$, or SCRPA, $E_c^\mathrm{SCRPA}$, to be added to the
Hartree-Fock $E^\mathrm{HF}$ kinetic, external, Hartree and exchange
contributions to the total energy, $E^\mathrm{RPA} = E^\mathrm{HF} +
E_c^\mathrm{RPA}$.  The correlation contribution can be calculated by
[see Eq.\ (8.111) in Ref.\ \cite{RingSchuck}]
\[
 E_c = - \sum_{\lambda > 0} \Omega_\lambda \sum_{ph} |\chi^\lambda_{hp}|^2
 ,
\]
but also by the expression [see Eq.\ (8.94b) in Ref.\ \cite{RingSchuck}]
\[
 E_c = \frac{1}{2} \sum_{\lambda > 0} \left( \Omega_\lambda^\mathrm{full} - \Omega_\lambda^\mathrm{TDA} \right)
 ,
\]
(the sums over $\lambda$ run only over the positive $\Omega_\lambda$ energies)
implying a difference between the excitation energies obtained by
solving the full RPA Eq.\ (\ref{RPA}), and excitation energies in the
Tamm-Dancoff approximation (TDA), obtained by neglecting the coupling
terms $B$ between the particle-hole and the hole-particle sectors of the
full matrix in the solution of the Eq.\ (\ref{RPA}).  The
two formulas gave the same results well within the accuracy quoted in
this work, and so provided a cross check over the validity of the
total-energy results.  The same formulas were also used to calculate
the total energy in the BSE approach. 

To perform the renormalized RPA calculation on helium we first
calculated the HF ground state and electronic structure energies and
wave functions $\epsilon^\mathrm{HF}_i, \phi^\mathrm{HF}_i(\vec{r})$
by solving the Hartree-Fock equations
\[
  H_\mathrm{H}(\vec{r}) \phi^\mathrm{HF}_i(\vec{r}) + \int d\vec{r}' \, 
     \Sigma_x(\vec{r},\vec{r}') \phi^\mathrm{HF}_i(\vec{r}')
  = \epsilon^\mathrm{HF}_i \phi^\mathrm{HF}_i(\vec{r})
  ,
\]
where $H_\mathrm{H}(\vec{r}) = - \partial_\vec{r}^2/2 +
v_\mathrm{ext}(r) + v_\mathrm{H}(\vec{r})$ is the Hartree Hamiltonian
and $\Sigma_x$ is the Fock exchange operator.  We did not rely on
pseudopotentials and rather use the full nuclear potential
$v_\mathrm{ext}(r) = - Z / r$ to reduce any source of inaccuracy in
our comparison to the exact result.  The HF calculation was carried out by
the \textsc{nwchem} package \cite{nwchem}.  With the HF electronic
structure we calculated the $\mathcal{S}$ matrix of the RPA equation
(\ref{SCRPA}) and then solved it to get the standard RPA excitations
(both singlet and triplet) energies $\Omega_\lambda$ and amplitudes
$\chi^\lambda$.  These are the excitations that we report in our
tables and figures as (standard) RPA or TDHF, and are also the
first-iteration result of an r-RPA calculation towards
self-consistency.  We then used the $\chi^\lambda_{hp}$ amplitudes to update
the occupation numbers [Eqs.\ (\ref{occ1}) and (\ref{occ2}), where
  $\lambda$ run over both singlet and triplet excitations] and energies
$\epsilon_k$ from Eq.\ (\ref{MFeq}), which are reinjected into the RPA
equation to be solved again for new $\chi^\lambda$ amplitudes.  The
procedure was repeated until self-consistency, 
(at most four cycles were enough to achieve the
$10^{-4}$ Ha accuracy we quote).  The r-RPA calculations were carried
out using a modified version of the \textsc{Fiesta} code
\cite{BlaseOlevano11,Jac15a}.  We used the \emph{d-aug}-cc-pV5Z
\cite{Dunning94} correlation-consistent Gaussian basis set with
angular momentum up to $l=5$ and including a double set of diffuse
orbitals.  This was the most converged basis set and the best
available to us.

\begin{table*}[t]
\begin{tabular}{crr|crr}
  \hline \hline
  \multicolumn{3}{c|}{cc-pV5Z} & \multicolumn{3}{c}{\textit{d-aug}-cc-pVQZ} \\
  \hline
 $n^S\!L$ & Full-CI & ICE-CI & $n^S\!L$ & Full-CI & ICE-CI\\
  \hline
 $1^1\!S$ & $-2.903151884$ & $-2.903151884$ & $1^1\!S$ & $-2.902536607$ & $-2.902536607$ \\
 $2^3\!S$ & $-2.041940640$ & $-2.041940640$ & $2^3\!S$ & $-2.174798591$ & $-2.174798592$ \\
 $2^1\!S$ & $-1.923273478$ & $-1.923273482$ & $2^1\!S$ & $-2.145020288$ & $-2.145020287$ \\
 $2^3\!P$ & $-1.714041381$ & $-1.714041383$ & $2^3\!P$ & $-2.130703422$ & $-2.130703422$ \\
 $2^1\!P$ & $-1.593255618$ & $-1.593255621$ & $2^1\!P$ & $-2.119799159$ & $-2.119799159$ \\
 $3^1\!S$ & $-0.588140506$ & $-0.588140506$ & $3^3\!S$ & $-2.063091342$ & $-2.063091342$ \\
 $3^3\!S$ & $-0.575726092$ & $-0.575726092$ & $3^1\!S$ & $-2.046569475$ & $-2.046576198$ \\
 $3^3\!P$ & $-0.390104384$ & $-0.390104386$ & $3^3\!D$ & $-1.920654679$ & $-1.920654679$ \\
 $3^1\!P$ & $-0.326412907$ & $-0.326412909$ & $3^1\!D$ & $-1.920163475$ & $-1.920163475$ \\
 \hline \hline
\end{tabular}
\caption{Helium excitation energies in atomic units (Ha), comparing
  Full-CI and ICE-CI results within the cc-pV5Z and the
  \textit{d-aug}-cc-pVQZ Gaussian basis sets.  The zero of energy is
  the full ionization onset, He$^{++} + 2e^-$.}
\label{full-ice-ci}
\end{table*}

\subsection{QMC}

We performed variational and diffusion quantum Monte Carlo (VMC and
DMC) calculations \cite{Ceperley80,Foulkes01} of the nonrelativistic
ground-state energy of an isolated all-electron helium atom with
infinite nuclear mass.  The \textsc{casino} code was used to perform
our calculations \cite{Needs10}. The ground-state wave function is
nodeless and hence the DMC algorithm is unbiased in the limit of zero
time step, infinite walker population, and sufficiently long
equilibration time.

We used a trial wave function of Slater-Jastrow form \cite{Foulkes01}:
\begin{equation} \Psi({\bf r}_1,{\bf r}_2)=\phi_{1s}^{\rm HF}(r_1)\,\phi_{1s}^{\rm HF}(r_2) \, \exp(J), \label{eq:sj_wf} \end{equation}
where the Jastrow exponent is \cite{Drummond04}
\begin{eqnarray}
J & = & \sum_l \alpha_l r_{12}^l(r_{12}-L_u)^3\Theta(L_u-r_{12})
\nonumber \\ & & {} +\sum_i\sum_m \beta_m r_i^m
(r_i-L_\chi)^3\Theta(L_\chi-r_i) \nonumber \\ & & {} + \sum_{l,m,n}
\gamma_{lmn} r_1^l r_2^m r_{12}^n (r_1-L_f)^3(r_2-L_f)^3 \nonumber
\\ & & \hspace{7em} {} \times
\Theta(L_f-r_1)\Theta(L_f-r_2), \end{eqnarray} where $\Theta$ is the
Heaviside function.  The electron orbital $\phi_{1s}^{\rm HF}$ in the
Slater part was calculated using Hartree-Fock theory and was
represented numerically on a radial spline grid, allowing the
electron-nucleus Kato cusp condition to be satisfied
\cite{Kato57,Pack66}.  The Jastrow exponent consisted of polynomial
electron-electron, electron-nucleus, and electron-electron-nucleus
terms, which were smoothly truncated at distances of $L_u=8$,
$L_\chi=8$, and $L_f=6$ Bohr, respectively \cite{Drummond04}.
Constraints were imposed on the parameters $\alpha_l$, $\beta_m$, and
$\gamma_{lmn}$ to enforce the electron-electron Kato cusp condition
and to avoid interfering with the electron-nucleus cusp condition; the
remaining parameters were optimized. The Jastrow factors used in the
great majority of QMC calculations are of this form or similar.  Since
the exact helium-atom wave function is a function solely of the
electron-nucleus and electron-electron distances, the helium atom is a
favorable case for our Jastrow exponent, which is a polynomial
expansion in these distances.  Free parameters in our trial wave
function were optimized by energy minimization \cite{Umrigar07}.  Our
wave function contained 42 free parameters, and optimization of the
wave function required about 32 core hours of computational effort.
The resulting VMC energy is $-2.90372220(7)$ Ha.  This is lower than
the VMC energy [$-2.903693(1)$ Ha] reported with a similar form of
Jastrow factor in Ref.\ \cite{Drummond04} due to the use of a
different optimization method.

In our DMC calculations we used time steps of 0.002 and 0.008
Ha$^{-1}$, with corresponding target populations of 1024 and 256
walkers.  The resulting DMC energies were extrapolated linearly to
zero time step and hence, simultaneously, to infinite population.  The
resulting DMC energy of a helium atom with infinite nuclear mass is
$-2.9037246(9)$ Ha.  The total cost of the two DMC calculations was
121 core hours.

The DMC result is within error bars of the exact energy, as shown in
Table \ref{groundstate}.  This is to be expected, as DMC is a
statistically exact method for helium.  However, a small difference
between the VMC result and the exact result can be seen.  This is due
to the finite extent and order of the polynomials in the wave
function, and the use of a finite number of random configurations in
the wave-function optimization.

\subsection{CI}

The configuration interaction (CI) results we report here are standard calculations for which we
invite the reader to refer to the specialized literature
\cite{HelgakerJorgensenOlsen}.  We were able to perform full-CI
\cite{HelgakerJorgensenOlsen} calculations for all Gaussian basis sets employed  here, except for the
\textit{d-aug}-cc-pV5Z which is fundamental to get the best
convergence for excited states.  This basis set is already too large
to allow a full-CI calculation, at least for the computing resources
available to us.  This is already an important indication of the
extent to which a given methodology, here CI, is able to achieve in
practice.  However, for \textit{d-aug}-cc-pV5Z we performed an 
iterative-configuration expansion configuration interaction (ICE-CI)
calculation \cite{ice-ci,orca2} and we checked, within the cc-pV5Z and the
\textit{d-aug}-cc-pVQZ basis sets, that ICE-CI provides results that
are indistinguishable from full-CI, in particular on the ground state
where the difference is $< 10^{-9}$ Ha, and remains generally at
$10^{-9}$ Ha for most excited states, except in one case, where the
difference was found to be $7 \cdot 10^{-6}$ Ha (see Table
\ref{full-ice-ci}), well beyond what can be considered the accuracy of
CI with respect to the exact solution.  
It would be unfair not to say that, beyond ICE-CI, there are several other methods able to provide near-full-CI energies, like CIPSI, CCSD, etc.
We again invite the reader to refer to the specialized literature \cite{HelgakerJorgensenOlsen,LoosJacquemin18,LoosJacquemin19,GarnironCaffarel17}
In the rest of the paper we
therefore quote these results (with a number of decimal digits equal
to or less than 6) as CI \textit{tout court}, irrespective of whether
they were obtained using full-CI or ICE-CI\@.  All CI calculation were
carried out using the publicly available \textsc{orca} code
\cite{orca,orca2}.

\subsection{HF+dRPA, \textit{G\hspace{-.1em}W}+dRPA, BSE}

The results and the details of many-body perturbation theory using the
Bethe-Salpeter equation have been already reported by some of us
in a previous publication \cite{LiOlevano17} to which we refer the
reader for both the theory and the parameters used in the calculation.  In
practice, the BSE equation is very similar to the standard RPA
equation [Eq.\ (\ref{RPA})].  The major differences are that the RPA
kernel $\bar{v}$, Eq.~(\ref{RPAkernel}), is replaced by a kernel
\[
 K^\mathrm{BSE}_{ijkl} = \langle ij | v | kl \rangle - \langle ij | W | lk \rangle
 \]
in the BSE equation, where the second term has been replaced by matrix
elements of the screened Coulomb interaction $W$, instead of the bare
Coulomb interaction $v$.  Another major difference is that the BSE
calculation is done on top of an already correlated $GW$ electronic
structure instead of the HF uncorrelated electronic structure
used in standard RPA (see Fig.~\ref{polarizabilities}).

In this work we add some other new results obtained using the dRPA approximation,
that is the direct RPA without exchange diagrams.
The difference between the two can be understood when looking at the
Feynman diagrams entering the irreducible polarizability $\widetilde{\Pi}$
(see Fig.~\ref{polarizabilities}): in the
dRPA polarizability only the particle-hole bubble (ring)
diagram enters, while in the full RPA both the ring bubble and also
the particle-hole exchange bubble diagram enter.  In both cases then
the irreducible polarizabilities $\widetilde{\Pi}$ are resummed up 
to infinity to get the reducible polarizability $\Pi$
(last line of Fig.~\ref{polarizabilities}). The dRPA is therefore a
lower level of approximation. Another often used name
for the dRPA is \textit{ring approximation}, with reference to the
diagrams taken into account.

We report new results for helium obtained using
this dRPA on top of both
Hartree-Fock and also $GW$ electronic structures.  So, the
particle-hole ring bubbles are calculated using electron and hole
Green's functions relying in one case on the HF electronic structure
(HF+dRPA), and in the other on the quasiparticle $GW$ ($GW$+dRPA,
see Fig.~\ref{polarizabilities}).
The comparison between the HF+dRPA with the RPA results and between
the $GW$+dRPA with the BSE result, will show us the effect of the
electron-hole interaction represented, in the first case by the bare
Coulomb interaction, and by the screened Coulomb interaction in the
second (Fig.~\ref{polarizabilities}).
Like in all the other cases, we used the same ingredients,
Gaussian basis sets (\textit{d-aug}-cc-pV5Z), and parameters of the
calculation so to allow the most faithful comparison between methods.
The calculations were carried out using again the \textsc{Fiesta} code,
switching off the electron-hole interaction term of the kernel.

\subsection{TDDFT}

TDDFT calculations  share a lot of similarities with the
standard RPA\@.  In TDDFT excitation energies and
amplitudes are calculated by solving also the RPA equations
[Eq.~(\ref{RPA})], which in chemistry are called the Casida equations
\cite{Casida95,Casida96}.  The differences with respect to standard RPA are that
(see Fig.~\ref{polarizabilities}): 1)
The DFT Kohn-Sham electronic structure is used instead of the HF
electronic structure to calculate the zero-order polarizability; 2) The
kernel $\bar{v}$, Eq.~(\ref{RPAkernel}), of the standard RPA equations
is replaced by a TDDFT kernel $f^\mathrm{TDDFT}$ given by
\begin{equation}
 f^\mathrm{TDDFT}_{ijkl} = \langle ij | v | kl \rangle + \langle ij |
 f_{xc} | kl \rangle
 . \label{TDDFTkernel}
\end{equation}
The first term is exactly the same in standard RPA, BSE and TDDFT kernels.
TDDFT replaces the second exchange term of RPA or the $W$ term of BSE,
by a direct term called exchange-correlation kernel, $f_{xc}$,
which is defined as the functional derivative of the
exchange-correlation potential with respect to density (the second
functional derivative of the exchange-correlation energy):
\[
  f_{xc}[\rho](x_1,x_2) = \frac{\delta v_{xc}[\rho](x_1)}{\delta
    \rho(x_2)} = \frac{\delta^2 E_{xc}[\rho]}{\delta \rho(x_2) \delta
    \rho(x_1)} .
\]
TDDFT is an in-principle-exact framework to calculate neutral excitation
energies and oscillator strengths.  However, the exact form of
$f_{xc}$ is in general unknown.  The latter in particular is in principle a dynamical quantity
depending on time and hence on frequency \cite{DelSoleReining03}. So one must resort to
approximations.  The adiabatic local-density approximation (ALDA or
TDLDA) is one of the most popular and consists of taking the
functional derivative of the DFT local-density approximation to the
exchange-correlation potential with respect to the density.  Here we
report calculations using this TDLDA approximation on top of both a
DFT-LDA Kohn-Sham electronic structure as well as the exact DFT
Kohn-Sham electronic structure.  The latter is reported in
Ref.\ \cite{SavinGonze98,UmrigarGonze94}, and was done using a
real-space-real-time code.  On the other hand, we carried out
DFT-LDA+TDLDA calculations using the \textsc{nwchem}
code relying once again on the same basis set and
calculation parameters as in all other calculations, in particular the
\textit{d-aug}-cc-pV5Z basis. 

Finally, we also report the results of a DFT-LDA+dRPA calculation,
that is using the dRPA approximation on top of a DFT LDA Kohn-Sham
electronic structure. This is equivalent to a TDDFT calculation
neglecting completely the exchange-correlation kernel, $f_{xc}=0$,
in Eq.~(\ref{TDDFTkernel}) \cite{BottiAndreani02}.
The comparison between DFT-LDA+dRPA and DFT-LDA+TDLDA results
shows the effect of the approximated exchange-correlation kernel 
$f_{xc}$.

\section{Results}

In this section we will compare the results provided by the different
methods, starting with the ground-state energy and then moving to
excitations.

\begin{table}[t]
 \begin{tabular}{ll}
  \hline \hline
  Method & Energy [Ha] \\
  \hline
  Noninteracting & $-4$ \\
  Hartree   & $-1.9517$ \\
  HF        & $-2.8616$ \\
  DFT-LDA   & $-2.8348$ \\
  DFT-GGA   & $-2.8929$ \\
  Exact-DFT \cite{UmrigarGonze94} & $-2.903724377034118$ \\
  RPA (TDHF)& $-2.9097$ \\
  r-RPA    & $-2.9085$ \\
  $GW$+BSE    & $-2.9080$ \\
  CI        & $-2.9032$ \\
  QMC-VMC (SJ)  & $-2.90372220(7)$ \\
  QMC-DMC   & $-2.9037246(9)$ \\
  \textbf{Exact} \cite{Schwartz06}     & $-2.903724377034119598311159245194404$ \\
  \hline
  \hline
 \end{tabular}
 \caption{Ground-state energy as calculated by different many-body
   approaches.  The zero of energy is set to the full ionization
   onset, He$^{++} + 2e^-$.  GGA refers to the PBE functional.  The
   exact-DFT result quoted from Ref.\ \cite{UmrigarGonze94} is
   calculated from the exact exchange-correlation potential obtained
   by reverse engineering from the exact Hylleraas solution.  So, its
   accuracy only reflects the accuracy of the Hylleraas solution that
   must be known in advance, in contrast to the accuracies of all
   other methods which are genuine and real.  We quote the exact-DFT
   result just to remind the reader of the scope of DFT, which should
   be the target of improved approximations.
}
 \label{groundstate}
\end{table}

\subsection{Ground-state}

In Table \ref{groundstate} we report the helium atom ground-state
energy (in atomic units [Hartree] and setting the zero of the energies to
the full ionization onset He$^{++} + 2e^-$) as calculated by all the
methods we have considered.  The exact result is quoted from the
Hylleraas-like Schwartz calculation \cite{Schwartz06}, which achieved
an accuracy of 35 decimal digits, further confirmed by later work
\cite{NakashimaNakatsuji07}.  The noninteracting energy
[Eq.\ (\ref{IP})] of 4.0 Ha presents a large error of $\sim$ 1 Ha
$\sim$ 30 eV from the exact result, showing how important are the
interactions between electrons and how crude is the
independent-particle approximation when calculating energies.  The
simplest many-body method, the Hartree-Fock (HF) theory, provides a total
energy of $-2.8616$ (our Gaussian \textit{d-aug}-cc-pV5Z HF
calculation converged up to $10^{-4}$ Ha), already an important reduction
of the error by almost two orders of magnitude down to $\sim$ 0.04 Ha
$\sim$ 1 eV\@.  This cannot at all be considered chemical accuracy
that requires an error one order of magnitude less.
Nevertheless, Hartree-Fock already provides a reasonable answer at
least for the total energy of the system, and we will see also
for the ionization potential. The difference between the
exact and the Hartree-Fock energy,
\begin{equation}
  E_c = E^\mathrm{Exact} - E^\mathrm{HF}
  , \label{Ec}
\end{equation}
is the more rigorous definition of the correlation energy.  In
helium, one of the few real systems where we know the exact total
energy, we can calculate exactly the correlation energy and see that
it is $E_c = -0.042 \, \mathrm{Ha} = -1.15 \, \mathrm{eV}$
(see Table~\ref{GSenergycomposition}), only 1.4\% of the total energy.

\begin{table}[t]
 \begin{tabular}{ll}
  \hline \hline
  Method & Correlation energy [Ha] \\
  \hline
  RPA (TDHF)     & $-0.0481$ \\
  r-RPA          & $-0.0469$ \\
  $GW$+BSE       & $-0.0464$ \\
  CI             & $-0.0416$ \\
  \textbf{Exact} & $-0.0421$ \\
  \hline
  \hline
 \end{tabular}
 \caption{Ground-state correlation energy for the different many-body approaches, 
   obtained by subtracting from the total ground-state energy (Table~\ref{groundstate})
   the Hartree-Fock energy calculated at the same \textit{d-aug}-cc-pV5Z Gaussian basis set
   (Table~\ref{groundstate} second line).
   The exact correlation energy, calculated by Eq.~(\ref{Ec}), is converged to the
   same accuracy of the Hartree-Fock \textit{d-aug}-cc-pV5Z calculation ($10^{-4}$ Ha.)
}
 \label{gscorrelation}
\end{table}

Next we analyze the DFT-LDA result which presents an error larger
(almost the double) than that of HF: 0.07 Ha $\sim$ 1.9 eV\@.  A
DFT generalized gradient approximation (GGA) \cite{LangrethMehl83,Becke88} calculation  
(we used the most popular PBE functional \cite{PerdewErnzerhof96}) reduces
the error below the HF one: 0.01 Ha $\sim$ 0.3 eV\@.  Notice that
these are errors of the approximation, LDA or GGA (PBE), not of DFT,
which is in principle an exact theory to calculate the total ground-state energy.
The latter is just an idealistic statement which is better to avoid.
Nevertheless, these
statements are important to identify the scope and the limits of a
theory, and orient the research to the real challenges within these
limits \cite{MedvedevLyssenko17}.  
Helium is one of the few cases where this statement is not purely
idealistic, and the ``exact DFT'' that mathematical theorems
guarantee to exist, can be really touched by hands.  Thanks to the
existence of the exact Hylleraas solution, the exact
exchange-correlation potential of DFT can be calculated by reverse
engineering \cite{UmrigarGonze94,VarsanoGuidoni14}.  The exact Hylleraas ground-state
wave functions allows the calculation of the exact electron density,
and from the latter we can calculate the only occupied DFT Kohn-Sham wave
function.  Knowing the exact Kohn-Sham occupied wave function and its
corresponding Kohn-Sham energy (equal to the exact ionization
potential also provided by the Hylleraas solution), the Kohn-Sham
equation can be inverted to provide the exact exchange-correlation
potential of DFT\@.  This is the potential plotted in Fig.\ 8 of
Ref.\ \cite{UmrigarGonze94}.  Using the exact exchange-correlation
(XC) potential we can run the exact DFT and, for example, calculate
the total ground-state energy (Table I of Ref.\ \cite{UmrigarGonze94}
reported in our Table \ref{groundstate}) which,
with no surprise, coincides with the exact Hylleraas energy.
So for helium exact DFT is something more than only an idealistic theory. 
We cannot predict anything not already provided by the
Hylleraas solution, but we can at least study the DFT methodology.
Unlike all other entries in Table \ref{groundstate}, the ``Exact-DFT''
line is there not to indicate the actual performances of DFT in general, but
just to show that an exact exchange-correlation potential exists
and is able to provide the exact ground-state total energy by a \textit{mono-determinantal} (Kohn-Sham) approach
(but not other quantities outside the scope of Kohn-Sham DFT), 
and it is thus meaningful to search for approximate functionals that try to be as close as possible
to the exact potential also in the general case \cite{MedvedevLyssenko17}.

Next in our table we have a bunch of approximations that improve with
respect to Hartree-Fock up to one order of magnitude ($\sim$ 0.004 Ha
$\sim$ 0.12 eV for the $GW$+BSE ground-state energy).  The standard
RPA (TDHF) result presents a consistent improvement with
respect to HF\@.  Then both our r-RPA and the $GW$+BSE result
improve almost by the same, non-negligible but small amount, with
respect to standard RPA\@.

\begin{table}[t]
 \begin{tabular}{lrrr}
  \hline \hline
  CI  & cc-pV$x$Z & \textit{d-aug}-cc-pV$x$Z & \\
  \hline
 TZ & $-2.900232$ & $-2.900608$ & \\
 QZ & $-2.902411$ & $-2.902537$ & \\
 5Z & $-2.903152$ & $-2.903202$ & \\
 Extrapolation & $-2.903878$ &  $-2.903840$ & \\
 \textbf{Exact}\cite{Korobov02} & & & $\mathbf{-2.903724}$ \\
  \hline \hline
 \end{tabular}
 \caption{CI ground-state energy calculation, Gaussian basis set convergence.}
 \label{convergence}
\end{table}

\begin{figure}[t]
 \includegraphics[width=\columnwidth]{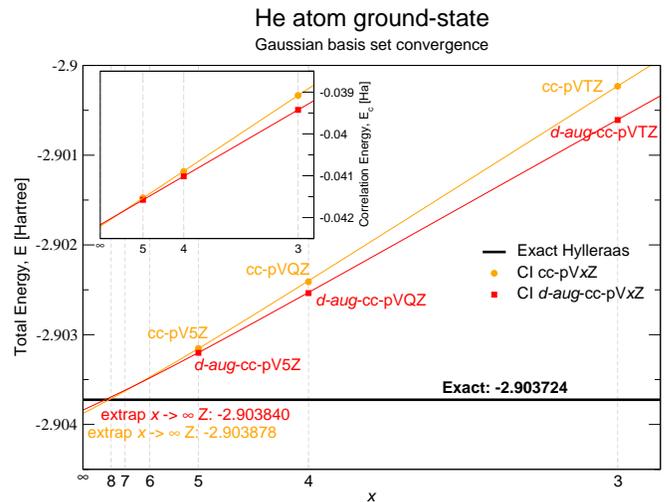}
 \caption{He atom ground-state $1^1\!S$ energy by CI calculations
   using increasing Gaussian basis sets.  In orange: calculations using
   the standard cc-pV$x$Z Gaussian basis set at increasing $x$ (orange
   circles) and their extrapolation to $x \to \infty$ (orange line).
   Red: calculations using the double augmented
   \textit{d-aug}-cc-pV$x$Z Gaussian basis set at increasing $x$
   (red squares) and their extrapolation to $x \to \infty$ (red
   line).  The Hartree-Fock and the correlation energies were
   separately fit to different formulas: the exponential function
   $E^\mathrm{HF}(x) = E^\mathrm{HF}_\infty + a e^{-bx}$ for
   Hartree-Fock, and a power law $E^c(x) = E^c_\infty + c x^{-3}$ for
   correlation (see inset).  Black line: Exact Hylleraas-like
   calculation ground-state energy \cite{Schwartz06}.  
   The zero of energy is set to the
   full ionization onset, He$^{++} + 2e^-$.  }
 \label{CI-GS}
\end{figure}

The first result that starts to be within the level of chemical
accuracy (usually set to 1 kcal/mol $\sim$ 0.0016 Ha $\sim$ 0.043 eV
\cite{HelgakerJorgensenOlsen}) is the CI result.
In Table \ref{groundstate} we quote our best
converged \textit{d-aug}-cc-pV5Z result, presenting an error
with respect to the exact result of $5 \cdot 10^{-4}$ Ha.
However, looking to Table
\ref{convergence} we can see that quadruple-Z Gaussian basis sets are
at the limits of chemical accuracy, and triple-Z, often the only
possibility for molecular calculations and by many considered as the
golden standard, are well outside.  Our study here tried also to
investigate to what extent the accuracy of CI can be improved.  Helium
is a very favorable case also for CI since the presence of only two
electrons limits the configurations to be taken in consideration
to singles and doubles only, with no need to include triples and beyond.
Nevertheless a CI calculation is to be done within a basis, here as in
most chemical calculations using a finite, incomplete Gaussian basis
set. This limits the accuracy of the calculation due to
two factors: 1) the number of configurations taken into account is
limited by the number of elements in the basis set; for example, in
the cc-pV5Z we have 55 basis elements, so that we can at best take
into account all the singles and doubles configurations out of 55
Hartree-Fock orbitals.  2) the chosen, localized or delocalized, basis
set can limit the representation of the exact wave functions; for
example, it can be very hard to represent the highest, almost free,
excited states by using a necessarily limited set of localized
Gaussians.  We tried to study to which extent the accuracy of CI with
respect to the previous issues can be pushed by applying a standard
\cite{CI_extrap0,CI_extrap,CI_extrap2} extrapolation technique 
over the $x$-tuple zeta Gaussian basis set series, towards the limit
$x \to \infty$. We fit the Hartree-Fock total
energies calculated at both the cc-pV$x$Z and the \textit{d-aug}-cc-pV$x$Z basis sets to the
exponential function $E^\mathrm{HF}(x) = E^\mathrm{HF}_\infty + a
e^{-bx}$, and separately the correlation energies to the power law
$E^c(x) = E^c_\infty + c x^{-3}$.  Figure \ref{CI-GS} is a plot of
this extrapolation technique compared to the exact Hylleraas energy
(in the inset the $x^{-3}$ linear extrapolation for the correlation
energy only).  We report in Table \ref{convergence} the extrapolated
values, $E^\mathrm{CI}_\infty = E^\mathrm{HF}_\infty + E^c_\infty$.
It can be seen that the extrapolation overshoots the exact result.  With
respect to the 5Z basis it provides a reduction of the error by a
factor 5, but it is unable to go below an error of $10^{-4}$ Ha. 
The 6Z basis would
present an error of the same magnitude of the extrapolated values, so
it would not be convenient to go for the extrapolation once
at the level, say, of 7Z or 8Z\@.  (Of course these values might be
system dependent.)  
Our analysis seems to show that a Gaussian CI
extrapolation technique towards the exact result 
is improved by
the \textit{augmentation} of the basis set. 
This implies that delocalized basis set elements,
that we will see are fundamental for the description of excited
states, are also important for an accurate description of the ground-state wave function and energy.

Quantum Monte Carlo is the most accurate among the many-body methods
studied here.  The VMC approach generally relies on a Slater-Jastrow
\textit{Ansatz} for the variational trial wave function, and this form
is used in nearly all QMC codes. Our VMC calculation achieves a random
error of only $7 \cdot 10^{-8}$ Ha, with the systematic error (bias
due to the restricted form of the trial wave function and the method
used to optimize the free parameters) being $2.18(7) \cdot 10^{-6}$
Ha.  Our DMC calculations using this VMC-optimized Slater-Jastrow wave
function achieved a statistical error of $9 \cdot 10^{-7}$ Ha, with no
evidence of systematic bias.  This demonstrates that QMC is
effectively able 
to achieve the experimental accuracy of Herzberg \cite{Herzberg58} used in the hystorical theory-experiment comparison of Pekeris \cite{Pekeris58}.
However, as noted earlier, helium is a very favorable case for QMC
because the ground-state wave function is nodeless; hence fixed-node
DMC is unbiased, i.e., one can obtain arbitrarily precise and accurate
DMC results by running for times as long as necessary.  The present
DMC calculation lasted 121 core hours.  The statistical error bar
falls off as the reciprocal of the square root of the computational
effort.  So the error bar can easily be reduced further, but the level
of precision achieved in Hylleraas calculations is completely
unachievable with QMC in practice, or would require a significantly
better trial wave function, together with an adaptation of a QMC code
for high-precision work.

The VMC accuracy achieved is 6 significant decimal digits, similar to
the accuracy of Kinoshita's 1957 Hylleraas-like calculation
\cite{Kinoshita57}. Fundamental to achieving this accuracy is the
cusp-like exponential factor $e^{-ks}$ present in the Kinoshita wave
function of Eq.\ (\ref{Psi}).  This factor is also present in the
standard VMC SJ \textit{Ansatz}.  So standard VMC can achieve
Kinoshita's, but not Frankowski and Pekeris's
\cite{FrankowskiPekeris66} accuracy of 14 significant decimal digits.
Achieving the latter might require the logarithmic factor $\ln(s)$ of
Eq.\ (\ref{Psiln}), which is absent in the standard VMC SJ
\textit{Ansatz} of multipurpose codes.  
One can easily implement such logarithmic behavior in the VMC
\textit{Ansatz}.  Knowing from the literature the best Hylleraas
result and the associated wave function, one could code an
\textit{Ansatz} modeled on the latter and possibly achieve the same
accuracy within VMC\@.  However, this is not the criterion we
have chosen in Table \ref{groundstate}, where the results are
deliberately obtained using ``standard'' methodology.  In any case,
the genuine QMC accuracies, compared to the rest of the methods that
do not require advance knowledge of the exact solution, are already
very impressive.

\begin{table}[t]
 \begin{tabular}{llll}
  \hline
  \hline
  Energy contribution & \multicolumn{1}{c}{HF} & Exact-DFT & \multicolumn{1}{c}{VMC} \\
  \hline
  Kinetic                 & $+2.8615$ & $+2.867082$ & $+2.90377(6)$ \\
  External (\textit{e-N}) & $-6.7489$ & $-6.753267$ & $-6.75332(6)$ \\
  Hartree              & $+2.0515$ & $+2.049137$ \\
  Exchange             & $-1.0257$ \\
  Exchange-Correlation & & $-1.066676$ \\
  Many-body (\textit{e-e}) & & & $+0.94585(5)$ \\
  \hline
  Total                & $-2.8616$ & $-2.903724$ & $-2.90372220(7)$\\
  Correlation          & $-0.0421$ \\
  \hline
  \textbf{Exact} \cite{Korobov02} & \multicolumn{3}{l}{$\mathbf{-2.903724377}$} \\
  \hline
  \hline
 \end{tabular}
 \caption{Different contributions to the ground-state energy in atomic
   units (Hartree), as calculated in our Gaussian
   \textit{d-aug}-cc-pV5Z HF calculation converged only up to
   $10^{-4}$ Ha (second column), and in the Exact-DFT calculation of
   Umrigar and Gonze \cite{UmrigarGonze94} accurate to the quoted digits
   (third column).  The 8th line reports the
   correlation energy, whose more rigorous definition is the
   difference between the exact and the Hartree-Fock energy.  The zero
   of energy is set to the full ionization onset, He$^{++} + 2e^-$.
   }
 \label{GSenergycomposition}
\end{table}

\subsection{Ground-state energy components}

It is also instructive to analyze the individual components of the 
total ground-state energy.
In Table \ref{GSenergycomposition} we report them for HF, QMC and once
again also for exact DFT for illustration purposes rather than
quantifying errors.  The total energy benefits from the
zero-variance principle (its error bar goes to zero as the trial wave
function is optimized) in both VMC and DMC \cite{Foulkes01}.  Hence
the total energy is much more precisely and a little more accurately
determined than its individual components.  It is possible to
calculate the exact energy components once the exact many-body wave
function is available from a Hylleraas calculation.  However we could
not find them in the literature.  We could anyway reconstruct what
should be the exact components, for example from the virial theorem
which in the case of Coulomb interacting systems says that the exact
kinetic energy must be minus the total energy.  By this argument we
can see that only VMC provides the expected behavior for the kinetic
energy, to within the error bars.  This is not the case for HF:
although the HF kinetic energy is virial with respect to its full HF
total energy, it does not coincide with the exact kinetic energy.  The
HF kinetic energy is the average value of the kinetic energy operator
over the HF single Slater determinant ground-state wave function, and
the latter is just an approximation to the exact many-body
ground-state wave function.

The same holds for exact DFT: the exact KS kinetic energy has nothing
to do with the exact kinetic energy $T$.  It is the kinetic energy
$T_s$ of the fictitious Kohn-Sham  independent-particle system,
i.e.\ the sum of the average kinetic energies of the Kohn-Sham
fictitious electrons.  In fact, the difference between the exact and
the Kohn-Sham kinetic energies, $T - T_s$, is included in the DFT
exchange-correlation energy $E_{xc}$ which, hence, contains also a part
of the real kinetic energy.  Here we can evaluate this part to be
$+0.036642$ Ha: this is almost the same magnitude (with change of sign) as
the correlation energy rigorously defined by Eq.~(\ref{Ec}),
$E_c = -0.0421$ (8th line in Table~\ref{GSenergycomposition}).
So, this kinetic contribution to the defined total exchange-correlation
energy $E_{xc}$ of DFT is not negligible at all with respect to the
correlation contribution.

The electron-nucleus external energy can be calculated once again
exactly (within the error bar) by VMC as the average of the external
potential local operator over the VMC wave function.  The external
energy can also be in principle calculated exactly within DFT: to
calculate this quantity the full many-body wave function is not
needed, just the electronic density, which is provided exactly with the
exact DFT\@.  The external energies of exact DFT and VMC coincide.
This of course is not the case in approximate (LDA, GGA, etc.)\ DFT\@.
On the other hand, HF provides only an approximate electronic density,
and so the external electron-nucleus energy provided by HF is only an
approximation for this component.

Ambiguities related to the definitions start to arise when looking at
the Hartree energy.  At the beginning this component was defined with
respect to a particular method, the Hartree or the Hartree-Fock
method.  These two methods already provide a different estimate for
the Hartree energy, due to the fact that the ground-state
wave functions and electronic densities are different.  
However the
Hartree energy can be defined as the classical component to the
many-body electron-electron energy:
\[
 E_\mathrm{H} = \int d^3r d^3r' \, \frac{\rho(\mathbf{r}) \rho(\mathbf{r}')}{|\mathbf{r}-\mathbf{r}'|}
 .
\]
With this definition, one can see that exact DFT also provides the
exact Hartree energy, again because the density is exact.  And we can
measure the error in this component within HF theory, which is related
to the error in the HF external energy.  In principle the charge
density and hence Hartree energy can be calculated within QMC, but in
practice QMC directly evaluates the many-body electron-electron
interaction energy (6th line of Table \ref{GSenergycomposition}).
Likewise for the exchange and correlation energies.  In fact, the
exchange energy can only be defined once the exchange operator, which
relies on single-particle wave functions, is defined, what is
meaningless in a QMC framework.  A meaningful exchange energy can only
be defined within the Hartree-Fock method and not within DFT even in
the exact case.  An exchange energy defined using the same HF shape
for the exchange operator but using Kohn-Sham (KS) wave functions,
i.e.\ the wave functions of the noninteracting KS electrons, has
not the same physical interpretation as the genuine Hartree-Fock exchange.  
In DFT normally one simply requires a full
exchange-correlation functional/potential that takes into account all
missing components together, including the kinetic energy not
accounted for by the Kohn-Sham kinetic energy $T_s$.  This is indeed
the case for exact DFT where the exchange-correlation energy exactly
provides the missing contribution (exchange plus correlation plus
residual non-Kohn-Sham kinetic energy) to achieve the exact total
ground-state energy (5th line in Table~\ref{GSenergycomposition}). 
Of course  DFT LDA, GGA, or other
approximations, should be evaluated for their error strictly done
on this quantity or on the density \cite{MedvedevLyssenko17}.

Finally from this table one can read off the exact value of the
correlation energy, rigorously defined with respect to the total exact 
and Hartree-Fock energies by Eq.~(\ref{Ec}), and so have
an estimate of its size and the only nonarbitrary and reliable evaluation
of how strongly or weakly correlated a many-body system is.
By comparing the correlation energy with the other contributions to the
total energy, one can appreciate how important correlations are in a 
given system, whether correlations are going to change qualitatively 
the picture or they are only a quantitative adjustment. 
In helium the correlation energy is more than
one order of magnitude less than all other components,
only a small fraction $< 5\%$ of them,
no matter how the other components are decomposed. 
So, the \textit{helium atom can be classified as a weakly correlated system}.

\begin{table}[t]
\begin{tabular}{cccc|cccc}
  \hline \hline
 $n^S\!L$ & BSE & CI & \textbf{Exact} & $n^S\!L$ & BSE & CI & \textbf{Exact}\\
 \hline
 \multicolumn{4}{c|}{atomic units [Ha]} & \multicolumn{4}{c}{electronvolt [eV]}\\
 $2^3\!S$ & 0.7271 & 0.7282 & \textbf{0.7285} & $2^3\!S$ & 19.786 & 19.815 & \textbf{19.824}\\
 $2^1\!S$ & 0.7676 & 0.7577 & \textbf{0.7578} & $2^1\!S$ & 20.888 & 20.618 & \textbf{20.621}\\
 $2^3\!P$ & 0.7724 & 0.7714 & \textbf{0.7706} & $2^3\!P$ & 21.018 & 20.991 & \textbf{20.969}\\
 $2^1\!P$ & 0.7894 & 0.7818 & \textbf{0.7799} & $2^1\!P$ & 21.480 & 21.274 & \textbf{21.222}\\
 $3^3\!S$ & 0.8427 & 0.8404 & \textbf{0.8350} & $3^3\!S$ & 22.930 & 22.868 & \textbf{22.722}\\
 $3^1\!S$ & 0.8637 & 0.8565 & \textbf{0.8425} & $3^1\!S$ & 23.502 & 23.307 & \textbf{22.926}\\
 $3^3\!P$ & 0.9514 & 0.9542 & \textbf{0.8456} & $3^3\!P$ & 25.890 & 25.965 & \textbf{23.010}\\
 $3^3\!D$ & 0.9645 & 0.9617 & \textbf{0.8481} & $3^3\!D$ & 26.247 & 26.169 & \textbf{23.078}\\
 $3^1\!D$ & 0.9663 & 0.9621 & \textbf{0.8481} & $3^1\!D$ & 26.294 & 26.180 & \textbf{23.078}\\
 $3^1\!P$ & 0.9928 & 0.9829 & \textbf{0.8486} & $3^1\!P$ & 27.015 & 26.746 & \textbf{23.092}\\
 \hline \hline
\end{tabular}
\caption{ He excitation energies in Hartree and eV, comparison between
  BSE and CI calculated at the \textit{d-aug}-cc-pV5Z basis against
  the exact \cite{KonoHattori84} result.  The zero of energy is set to the He ground state
  $1^1\!S$.  }
\label{exact-ci-bse}
\end{table}

\begin{figure}[t]
 \includegraphics[width=\columnwidth]{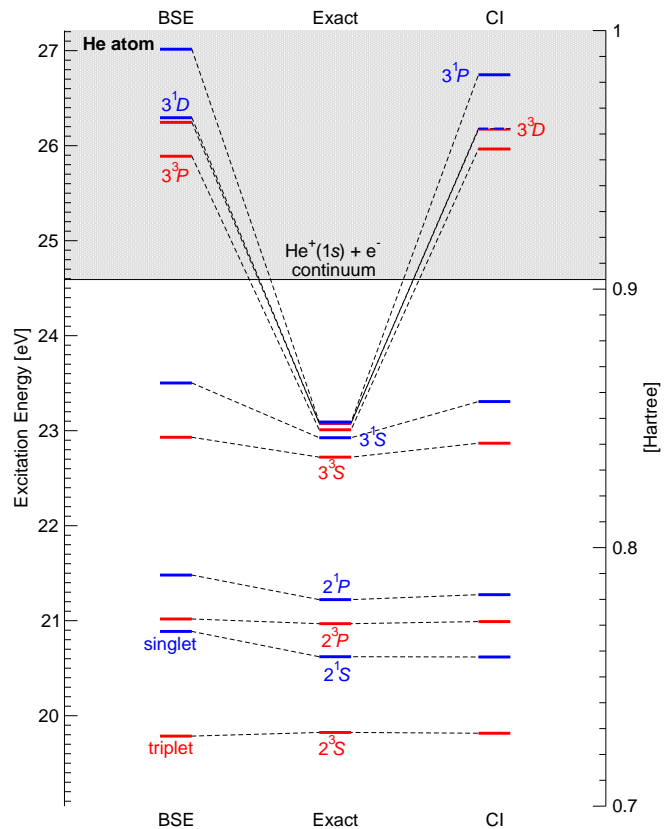}
 \caption{ He excitation energies in Hartree and eV, comparing
   BSE and CI results calculated within the \textit{d-aug}-cc-pV5Z basis
   against the exact \cite{KonoHattori84} result.  The zero of energy is set to the He
   ground state $1^1\!S$.  }
\label{exact-ci-bse-fig}
\end{figure}

\begin{table*}[t]
\begin{tabular}{cccccccccc}
 \hline \hline
 $n^S\!L$ & \parbox[c]{.095\textwidth}{\textbf{Exact}} & \parbox[c]{.095\textwidth}{CI} & \parbox[c]{.095\textwidth}{$GW$\\+\\BSE} & \parbox[c]{.095\textwidth}{TDHF\\(RPA)} & \parbox[c]{.095\textwidth}{HF\\+\\dRPA} & \parbox[c]{.095\textwidth}{$GW$\\+\\dRPA} & \parbox[c]{.095\textwidth}{DFT-LDA\\+\\dRPA} & \parbox[c][3.5em][c]{.095\textwidth}{DFT-LDA\\+\\TDLDA} & \parbox[c]{.095\textwidth}{Exact-DFT\\+\\TDLDA} \\
 \hline
 \multicolumn{9}{c}{atomic units [Ha]} \\
 $2^3\!S$ & \textbf{0.7285} & 0.7282 & 0.7271 & 0.7237 & 0.9396 & 0.9289 & 0.5826 & 0.5792 & 0.7351 \\
 $2^1\!S$ & \textbf{0.7578} & 0.7577 & 0.7676 & 0.7759 & 0.9414 & 0.9307 & 0.5882 & 0.5853 & 0.7678 \\
 $2^3\!P$ & \textbf{0.7706} & 0.7714 & 0.7724 & 0.7806 & 1.0136 & 1.0020 & 0.6381 & 0.6337 & 0.7698\\
 $2^1\!P$ & \textbf{0.7799} & 0.7818 & 0.7894 & 0.7997 & 1.0157 & 1.0041 & 0.6437 & 0.6340 & 0.7764 \\
 $3^3\!S$ & \textbf{0.8350} & 0.8404 & 0.8427 & 0.8499 & 1.0574 & 1.0444 & 0.6693 & 0.6575 & 0.8368 \\
 $3^1\!S$ & \textbf{0.8425} & 0.8565 & 0.8637 & 0.8732 & 1.0774 & 1.0644 & 0.7002 & 0.6872 & 0.8461 \\
 IP       & \textbf{0.9037} & 0.9179 & 0.9075 & 0.9179 & 0.9179 & 0.9075 & 0.5704 & 0.5704 & 0.9037 \\
 \hline
 \multicolumn{9}{c}{electronvolt [eV]} \\
 $2^3\!S$ & \textbf{19.824} & 19.815 & 19.786 & 19.692 & 25.569 & 25.276 & 15.853 & 15.760 & 20.003 \\
 $2^1\!S$ & \textbf{20.621} & 20.618 & 20.888 & 21.115 & 25.618 & 25.324 & 16.007 & 15.928 & 20.893 \\
 $2^3\!P$ & \textbf{20.969} & 20.991 & 21.018 & 21.242 & 27.581 & 27.266 & 17.363 & 17.244 & 20.947 \\
 $2^1\!P$ & \textbf{21.222} & 21.274 & 21.480 & 21.762 & 27.639 & 27.323 & 17.515 & 17.251 & 21.127 \\
 $3^3\!S$ & \textbf{22.722} & 22.868 & 22.930 & 23.128 & 28.773 & 28.421 & 18.214 & 17.891 & 22.770 \\
 $3^1\!S$ & \textbf{22.926} & 23.307 & 23.502 & 23.762 & 29.317 & 28.963 & 19.054 & 18.701 & 23.024 \\
 IP       & \textbf{24.591} & 24.979 & 24.696 & 24.979 & 24.979 & 24.696 & 15.522 & 15.522 & 24.591 \\
 \hline \hline
\end{tabular}
\caption{ He excitation energies in atomic units (Hartree, top) and
  electronvolt (eV, bottom), comparison between different methods.
  The zero of energy is set to the He ground state $1^1\!S$.  
  The exact DFT + TDLDA result is taken from Ref.~\cite{PetersilkaBurke00}
  and it is the only one in the table calculated without using the Gaussian
  basis set.
  The last line reports the ionization potential (IP), i.e.\ the first 
  ionization onset He$^+(1s) + e^-$, obtained from the last-occupied energy,
  $\mathrm{IP} = -\epsilon_{1s}$, of the (depending on the methodology)
  HF, $GW$, DFT LDA or exact electronic structures.
  Notice that for DFT LDA based calculations we could have used the
  $\mathrm{IP^{DFT-LDA}_{\Delta SCF}} = 0.8931$ Ha
  calculated by the $\Delta$SCF method, providing a fully bound Rydberg series,
  although severely red-shifted.
}
\label{excitations}
\end{table*}

\subsection{Excitations}

We now start to analyze excited states, starting from the comparison of CI
and exact results (Table \ref{exact-ci-bse}).  The first three CI
excitations are still within chemical accuracy from the exact result.
The agreement is still acceptable, within 0.5 eV, for the next three
excitations.  However, starting from $3^3\!P$ the error jumps to 3 eV
and more.  These states are also provided as unbound since the
ionization potential is set to $0.9037$ Ha.  This degradation is
evidently a finite-basis effect.  The lowest excited states are more
localized and require few Gaussians to be represented accurately.
Higher states get more and more delocalized and, consequently, require
larger (more diffuse) basis sets.  In particular we have found it is
essential to use augmented Gaussian basis set to describe even the
lowest excited states.  Looking at Table \ref{full-ice-ci} one can see
that the cc-pV5Z basis presents an error of more than 3 eV already on
the first excited state $2^3\!S$.  
This problem could be mitigated in large molecules because of the effect of basis elements sharing, i.e.\ the fact that each atom profits from the basis functions on its many neighbors.
However,
states towards the continuum of
hydrogenic He$^+$(1$s$) plus a free electron would require better
adapted bases, e.g., plane waves.  The states quoted in Table
\ref{exact-ci-bse} (see also Table \ref{full-ice-ci} for reference of
convergence) are the only ones that could be unambiguously identified,
though already in the unbound part of the spectrum.

The BSE approach is at the limit of chemical accuracy only for the
first excited state $2^3\!S$ and generally presents a larger error
than CI\@.  Very importantly, we observe the same trend as in CI, with
the characteristic breakdown at the level of the $3^3\!P$ excitation.
From that point on we observe a large error of both CI and BSE, but
the two methods are close to each other.  The worsening is evidently due
to basis-set incompleteness in both methods. 
The CI error can be regarded as mostly due to the incompleteness of the basis-set.
With this assumption, we can evaluate the error due to the approximations
done in the BSE formalism, independently from the basis set incompleteness error, 
by comparing directly BSE and CI results at the same basis set.
We see that this BSE formalism error is no more than 0.2~eV, 
an error that allows us to describe the main physics of a system.

In Ref.\ \cite{LiOlevano17} some of us already analyzed the results of
$GW$+BSE in comparison to RPA (TDHF)\@.  We now analyze
the results one can obtain from a dRPA (ring) approximation on top of
the HF or the $GW$ quasiparticle electronic structure.  
It can be seen in Table~\ref{excitations}
that the excitation energy is strongly overestimated in both approaches,
a nonrigid shift of 5--7 eV, and a slightly larger one with HF+dRPA\@.  The
difference between $GW$+BSE and $GW$+dRPA is a term that introduces
electron-hole (excitonic) screened interaction effects.  This also holds for
RPA (TDHF) and HF+dRPA, with the difference that we start from
uncorrelated Hartree-Fock energies and the electron-hole interaction
is unscreened.  One can see that this electron-hole interaction term
is very important at least in this isolated system, like it has
been found to be important in large band-gap insulators \cite{SottileReining07} and in molecules \cite{Jac15a,RangelRignanese11}.

\begin{table*}
\begin{tabular}{cccccccc}
  \hline \hline
   & \parbox[c]{.1\textwidth}{\textbf{Exact}} & \parbox[c]{.1\textwidth}{$GW$\\+\\BSE} & \parbox[c]{.1\textwidth}{TDHF\\(RPA)} & \parbox[c]{.1\textwidth}{HF\\+\\dRPA} & \parbox[c]{.1\textwidth}{$GW$\\+\\dRPA} & \parbox[c]{.1\textwidth}{DFT-LDA\\+\\dRPA} & \parbox[c][3.5em][c]{.1\textwidth}{DFT-LDA\\+\\TDLDA} \\ 
  \hline
  $f_{1^1\!S \to 2^1\!P}$ & \textbf{0.27616} & 0.2763 & 0.2916 & 0.1011 & 0.0996 & 0.1476 & 0.1848 \\ 
  \hline \hline
\end{tabular}

\caption{
   Helium atom first dipole-allowed $^1\!S \to 2^1\!P$ transition oscillator strength $f_{1^1\!S \to 2^1\!P}$.
   }
 \label{oscillatorstrength}
\end{table*}

In helium we know that the distance between the first ionization level
and the full ionization is exactly 2 Ha.  This is trivially given by
the solution of the Schr\"odinger equation for the system He$^+$, which
is a one-electron hydrogenic atom with $Z=2$.  So, the ionization
potential could be obtained by subtracting this value of 2 Ha from the
ground-state energy. However, in general, for systems with more than two
electrons, this information is not available.  We can then use
Koopmans' theorem: the last occupied HF eigenvalue, and more so the
corresponding $GW$ quasiparticle energy, can be interpreted as minus
the ionization potential (IP) of the atom.  This is $\mathrm{IP^{HF}}
= 0.9179$ Ha for HF and $\mathrm{IP^{GW}} = 0.9075$ for $GW$, against
the exact $\mathrm{IP^{exact}} = 0.9037$ Ha.  Referring to these values,
one can conclude that the excitation spectra of both HF+dRPA and
$GW$+dRPA are unbound (even the first excitation lie above the IP), in
contrast to the exact excitation spectrum, which presents a whole Rydberg
series below the ionization onset.  From this point one can see the
importance of running a BSE (or a TDHF / RPA) calculation using a kernel
that contains the electron-hole interaction (exchange term) beyond the
direct term of the simpler dRPA\@.

\begin{figure}[b]
 \includegraphics[width=\columnwidth]{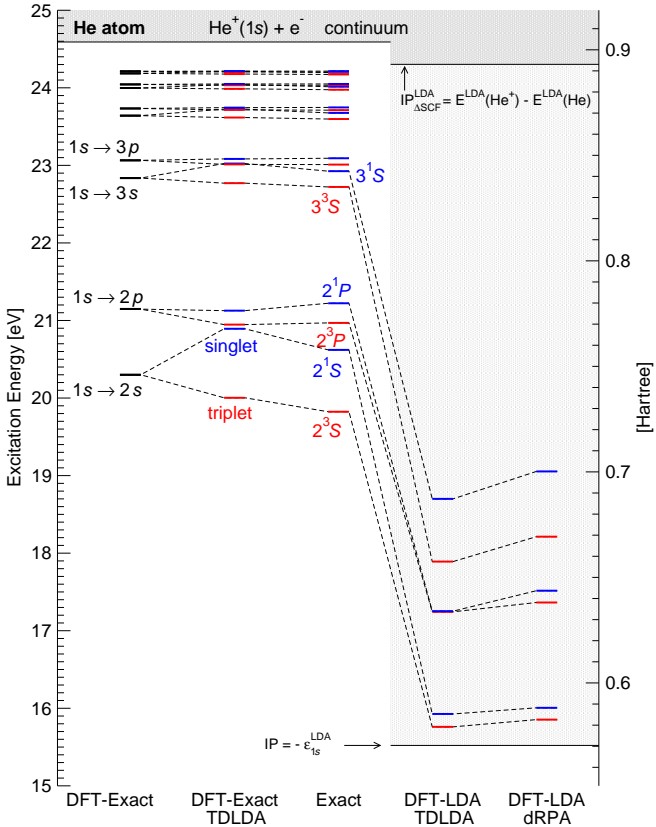}
 \caption{ He excitation energies in Hartree and eV\@.
    The zero of energy is set to the He
   ground state $1^1\!S$. From the left:
   exact-DFT \cite{UmrigarGonze98,SavinGonze98}; TDLDA on top of exact-DFT \cite{PetersilkaBurke00}; 
   exact spectrum \cite{KonoHattori84}; 
   TDLDA on top
   of DFT-LDA; dRPA on top of DFT-LDA\@.  The DFT-LDA spectra are
   calculated at the \textit{d-aug}-cc-pV5Z basis, while the exact-DFT
   are real-space calculations. 
   Notice that for DFT LDA based calculations we have used as onset of the
   continuum the $\mathrm{IP^{DFT-LDA}} = - \epsilon_{1s}^\mathrm{DFT-LDA} = 0.5704$ Ha,
   but we could have better used the $\mathrm{IP^{DFT-LDA}_{\Delta SCF}} = 0.8931$ Ha
   calculated by the $\Delta$SCF method.
   In the latter case we would have found a fully bound Rydberg series,
   and even severely red-shifted and overbound.
}
   \label{TDLDA}
\end{figure}

In DFT the last occupied Kohn-Sham eigenvalue is the only one that can
be physically interpreted as minus the ionization potential, that is
the energy to strip an electron from the system \cite{PerdewBalduz82,LevySahni84,AlmbladhVonBarth85}.  
In exact DFT the
last eigenvalue exactly coincides with minus the ionization potential.
In Ref.\ \cite{UmrigarGonze94} that value was taken from the exact
Hylleraas calculation and imposed for the inversion of the Kohn-Sham
equation.  In approximate DFT-LDA the last occupied Kohn-Sham
eigenvalue is supposed to give an approximate ionization potential in
order to estimate where the onset of the continuum of excitation
occurs.  This gives us $\mathrm{IP^{DFT-LDA}} = 0.5704$ Ha.  With
respect to this value it turns out that the DFT-LDA + dRPA spectrum is
also fully unbound (see Table~\ref{excitations} and Fig.~\ref{TDLDA}).
The same for a DFT-LDA + TDLDA spectrum.  There
is no Rydberg series before the onset of the continuum in DFT-LDA both with
dRPA and TDLDA\@.  Notice that if we had used the information that
the hydrogenic 1-electron helium ground-state energy is exactly 2~Ha
and calculated the IP  as the 2-electron helium DFT-LDA ground-state energy
(from Table~\ref{groundstate}) minus these 2~Ha, getting $\mathrm{IP} = 0.83$ Ha,
then we would have found a bound Rydberg series, although severely
red-shifted.  An always available and more convenient choice of the
ionization potential could have been obtained by taking the difference
between the DFT-LDA ground-state energies of the 2-electron and the
1-electron atoms, what is called the $\Delta$SCF method \cite{JonesGunnarsson89}.
Even though the 1-electron DFT LDA calculation is the most critically
affected by the self-interaction problem and error (that anyway our 1-electron DFT LDA 
calculation quantified to just only 0.06 Ha), by cancellation of errors with the
2-electron calculation a better result can be obtained:  
$\mathrm{IP^{DFT-LDA}_{\Delta SCF}} = 0.8931$ Ha.
So, we argue that in atoms, finding or not an unbound Rydberg series in
TDLDA (or dRPA) on top of DFT LDA calculations
depends, to a large extent, on the choice of how the IP has been calculated.

\begin{figure*}
  \includegraphics[width=\columnwidth]{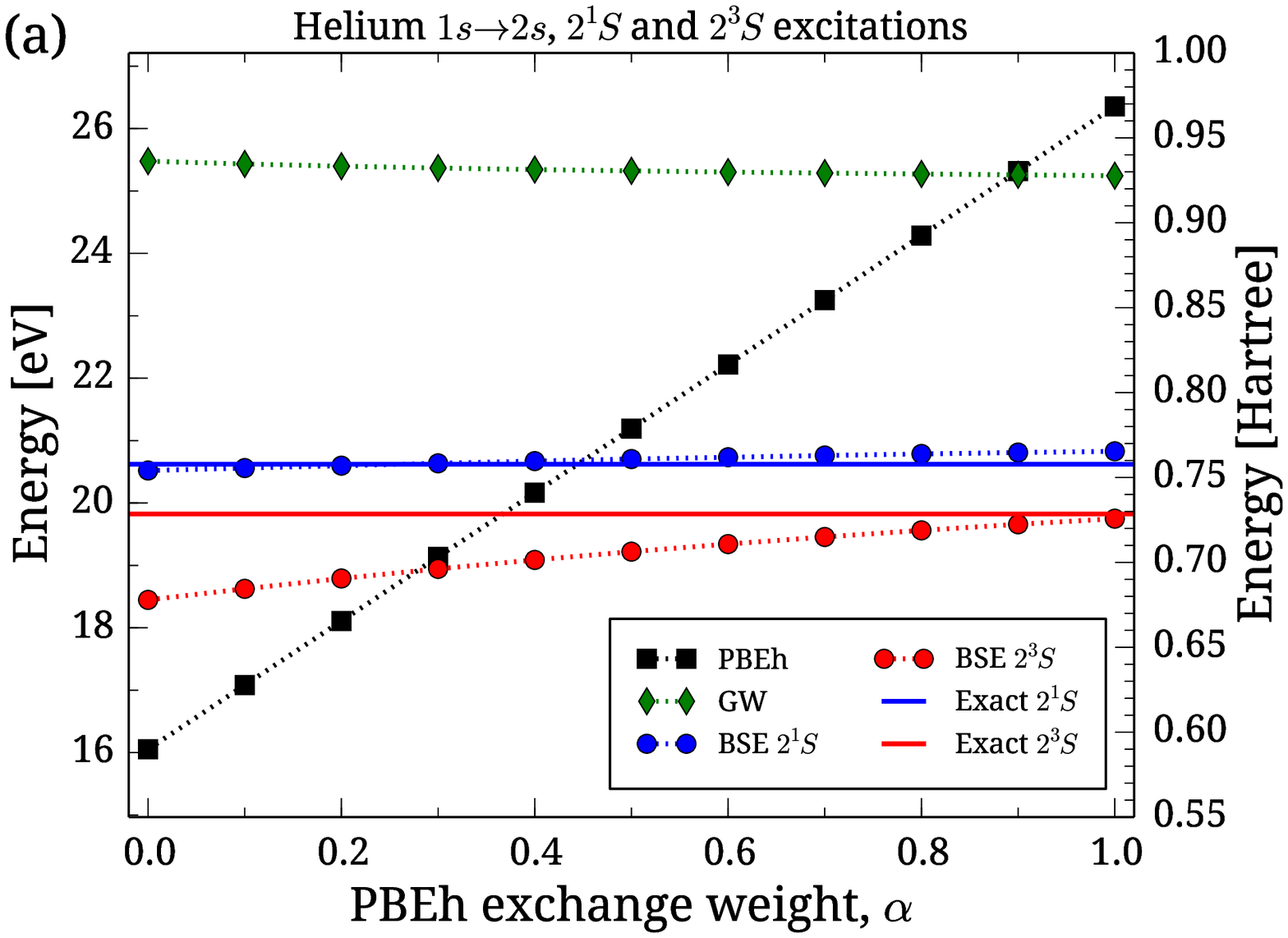} \hfil
  \includegraphics[width=\columnwidth]{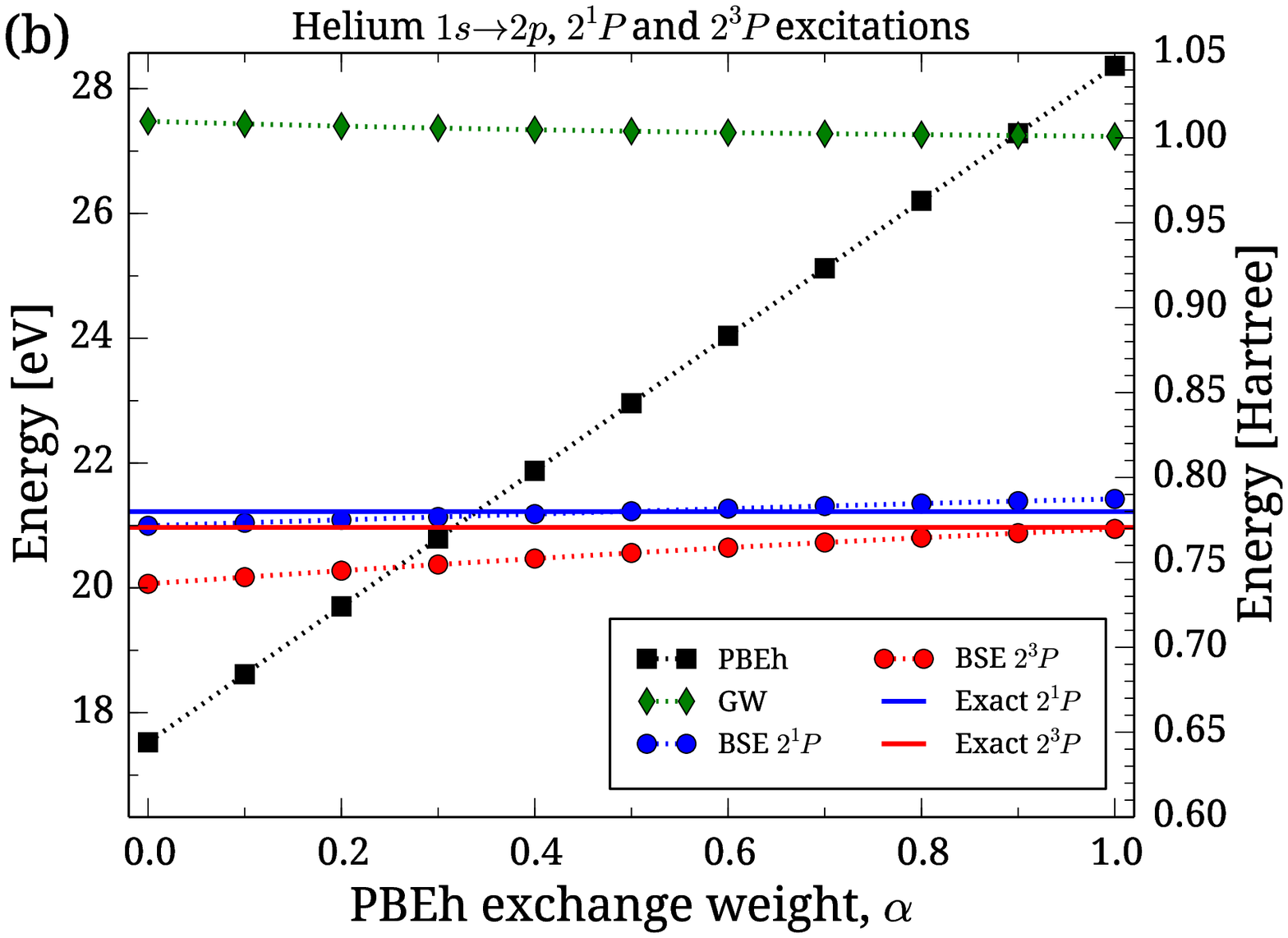}\\
  \includegraphics[width=\columnwidth]{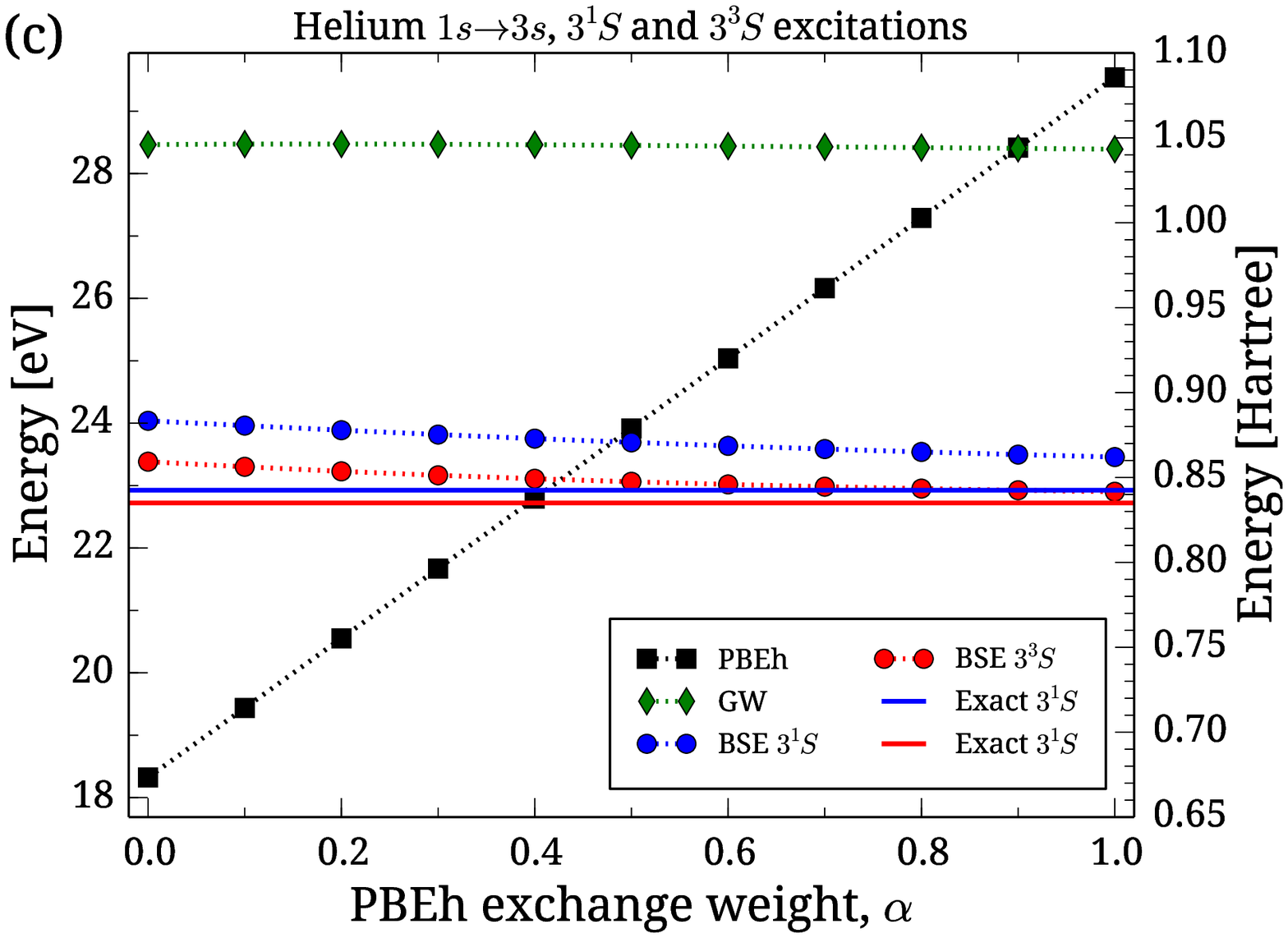} \hfil
  \includegraphics[width=\columnwidth]{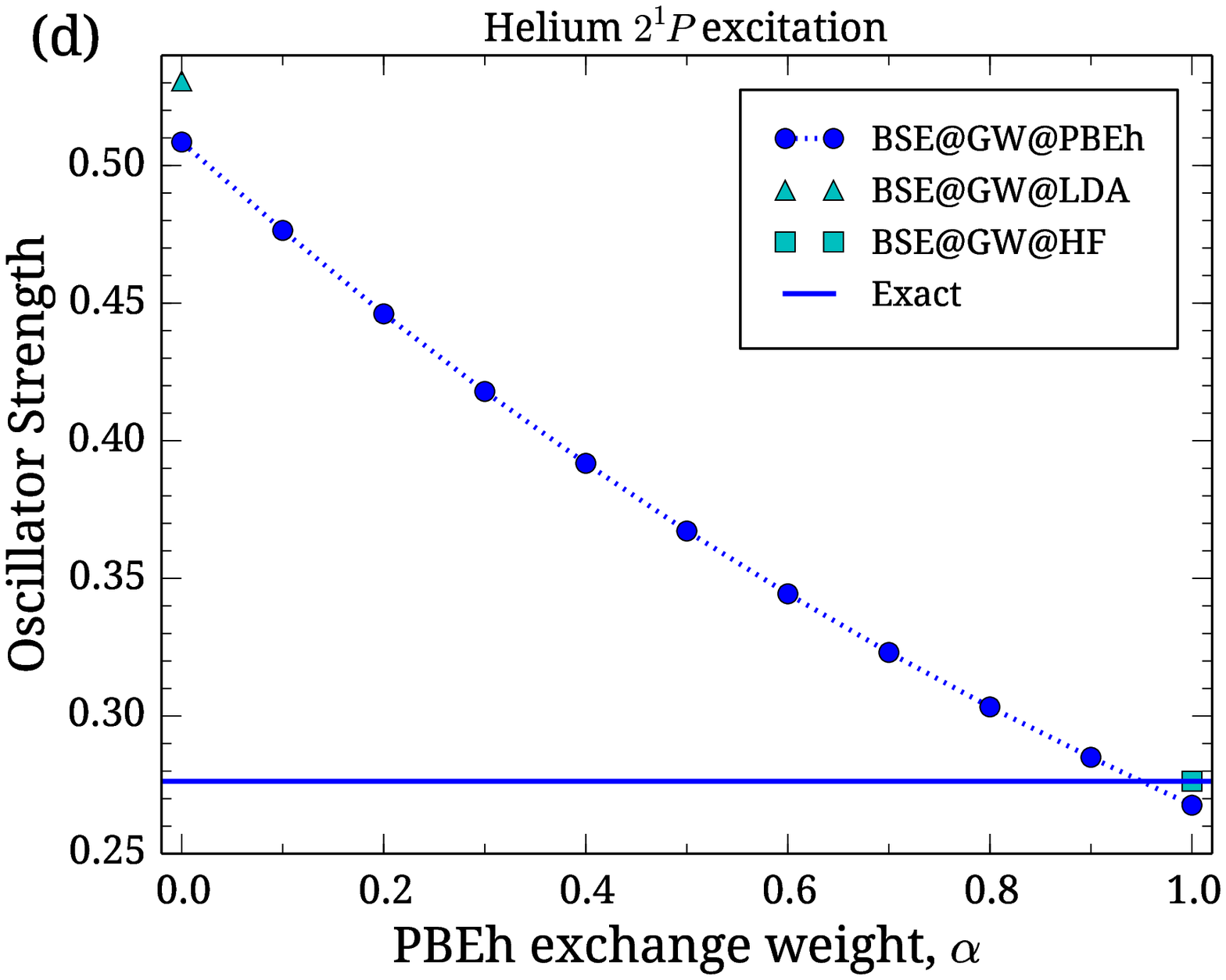}
  \caption{
  Starting point dependence, with respect to the PBEh exchange weight $\alpha$, of $GW$ and BSE results. $\alpha=0$ coincides with the original PBE \cite{PerdewErnzerhof96} functional, while $\alpha=1$ represents a full HF exchange plus the correlation contained in the PBE functional. 
  a-c) The $GW$ (green diamonds) and PBEh (black squares) gaps and the singlet (blue circles) and triplet (red circles) BSE excitation energies for a) $1s \to 2s$, b) $1s \to 2p$, and c) $1s \to 3s$, respectively. The corresponding exact singlet (blue solid line) and triplet  (red solid line) excitation energies are reported from Ref.~\cite{KonoHattori84}. 
  d) The $1^1\!S \to 2^1\!P$ BSE transition oscillator strength (blue circles).
  We report also the BSE transition oscillator strength obtained starting from pure HF (cyan square) and starting from pure DFT LDA (cyan triangle).
  The exact $1^1\!S \to 2^1\!P$ transition oscillator strength (solid line) is reported from Ref.~\cite{KonoHattori84}.  
 }
 \label{SP}
\end{figure*}

The use of the exact DFT Kohn-Sham spectrum \cite{UmrigarGonze98,SavinGonze98}
(Fig.~\ref{TDLDA} left side), for which the last occupied Kohn-Sham energy
provides the exact ionization potential \cite{PerdewBalduz82,LevySahni84,AlmbladhVonBarth85}, 
allows us to recover a bound Rydberg series in good agreement with the exact result. 
Indeed, an approximate TDLDA calculation done on top of exact DFT 
\cite{PetersilkaBurke00} is not any more affected by the two drawbacks
of the TDLDA calculation done on top of approximate DFT LDA, i.e.\ both the
unboundness of the entire spectrum due to the misplaced ionization potential,
and also the 3$\sim$5 eV severe shift of all excitations measured with
respect to the ground-state energy (see Fig.~\ref{TDLDA} and Table~\ref{excitations}).
Notice that, as is commonly done in solids, one can simulate this correction by applying a scissor operator to the DFT-LDA KS eigenvalue spectrum.
The LDA KS HOMO-LUMO gap of 15.853 eV has to be brought not to the exact HOMO-LUMO gap = IP - EA (electron affinity) of $\sim$25 eV, but rather to the exact ``optical gap'', i.e.\ at the 19.8 eV of the first excitation $2^3\!S$, or better at an average level of 20.2 eV between the singlet and triplet $2S$ excitations.
A scissor operator rigid shift of $4$--$\sim 4.4$ eV would better situate the DFT-LDA+TDLDA excitation spectrum.

To conclude this section we analyze the excitation oscillator
strengths (Table \ref{oscillatorstrength}).  This is a quantity
directly related to the quality of the wave functions.  By checking
oscillator strengths the different methodologies are evaluated with
respect to the quality of the wave functions, independently from
energies.  We note the good performances of BSE, but also of RPA,
against the unsatisfactory results of $GW$+dRPA and of HF+dRPA\@.  Like for
the excitation energies, both the unscreened kernel of RPA and the
screened one of BSE are fundamental to achieve good oscillator
strengths. 

\subsection{\textit{G\hspace{-.1em}W} and BSE starting point dependence}

\begin{figure}
 \includegraphics[width=\columnwidth]{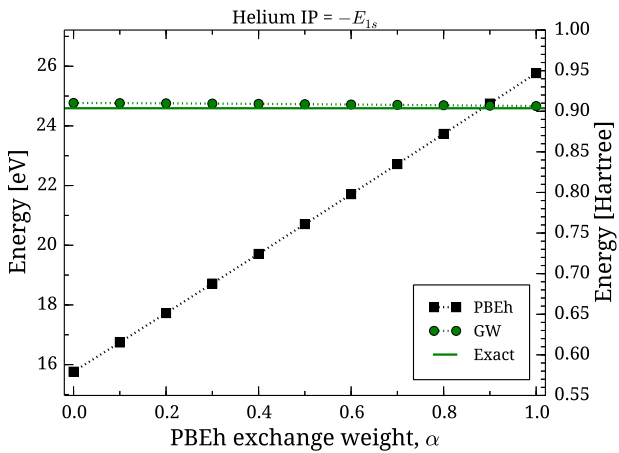}
 \caption{
  Starting point dependence, with respect to the PBEh exchange weight $\alpha$, of the ionization potential $\textrm{IP} = -E_{1s}$ as calculated in PBEh (black squares) and in the $GW$ approximation (green circles), compared to the exact \cite{KonoHattori84} value (green solid line).
 }
 \label{ionizationpotential}
\end{figure}

All the previously quoted results with $GW$ and BSE   have been calculated
starting from Hartree-Fock.  This is the approach of the origins
\cite{StrinatiHanke80,StrinatiHanke82,HankeSham79}
and it also looks to us more significative for a comparison
with TDHF and quantum chemistry methods like CI\@.  A dependence on the
starting point for $GW$ and BSE calculations should be expected,
although in this work we have performed a partial self-consistent $GW$ concerning only the energies.  In this section we will analyze the
dependence of both $GW$ and BSE results with respect to the starting point.  We
have chosen the hybrid DFT/HF PBEh approach \cite{AtallaScheffler13} with variable exchange
weight $\alpha$ \cite{Ernzerhof97} because this will allow to explore a full
range of situations.  From pure DFT-PBE at $\alpha=0$, to a HF
approach including correlation in the form of the local potential
associated to the PBE DFT functional at $\alpha=1$.  The results are
reported in Fig.\ \ref{SP}(a) for the $2S$ excitation energies (both
singlet and triplet), Fig.\ \ref{SP}(b) for the $2P$, and \ref{SP}(c) for the
$3S$.  It is quite surprising to see that in all the cases the value
of $\alpha$, that is the starting point, is little affecting the $GW$
HOMO-LUMO+n gaps, although PBEh gaps are strongly affected.  This is
also what we observe if we consider the ionization potential (IP),
equal to minus the energy of the HOMO $1s$ state
(Fig.\ \ref{ionizationpotential}).  However we point out again that we
performed a self-consistency on the $GW$ energies.  Wave functions on
the other hand are kept at the level of PBEh, and these can have a
more important effect on the matrix elements of the BSE kernel, and
consequently also on the BSE eigenvalues.  We observe such an
effect in Figs.\ \ref{SP}(a-c), in particular more on
the triplet states, while singlet states seem to follow the trend of $GW$ gaps. 
A much more important effect is to be expected on oscillator
strengths since the latter are only sensitive to wave functions. 
This is indeed what we observe in Fig.\ \ref{SP}(d) for the oscillator
strength of the transition $1^1\!S \to 2^1\!P$, varying in a broad range,
from $f=0.51$ at $\alpha=0$, to $f=0.27$ at $\alpha=1$

In conclusion, if the value of
$\alpha$ and the starting point seems to affect little the result of the
$GW$ gaps, and in part also  the energy of singlet excitations, a
choice of an $\alpha$ close to 1 seems to provide results more in
agreement with the exact calculation. This in particular for the oscillator
strength but also for the energy of triplet states, and finally also for
the ionization potential.  This seems to indicate that HF is the best
starting point for many-body perturbation theory calculations, at
least in the case of the helium atom and probably also of other isolated
systems.

We also report on a $GW$+BSE calculation starting from DFT LDA:
the results are close to the ones starting from DFT PBE 
(PBEh functional at $\alpha=0$).
When starting from DFT LDA, excitation energies are 0.2$\sim$0.3 eV larger
than when starting from DFT PBE,
like also the oscillator strength $f_{1^1\!S \to 2^1\!P}$, larger by 0.03
(see Fig.~\ref{SP}(d)).

Finally, we would like to compare our results with the available literature.
To the best of our knowledge, on helium atom there are no BSE calculations,
only $GW$ calculations of the ionization potential, $\textrm{IP} = -E_{1s}$, 
and these are one iteration $G_0W_0$
calculations starting from DFT LDA \cite{MorrisGodby07} or PBE \cite{VanSettenRinke15}.
Without performing any self-consistency, our fully dynamical contour-deformation
$G_0W_0$ calculation at the  \textit{d-aug}-cc-pV5Z Gaussian basis set provide an
ionization potential of 23.57 eV when starting from LDA, and 23.40 eV when starting from PBE\@. 
When starting from PBE, the best result Van Setten \textit{et al.}\ \cite{VanSettenRinke15}
have obtained is 23.48 eV, either using the codes \textsc{FHI-aims} (their analytic continuation
16 parameters Pad\'e approximant, P16 result) or \textsc{TURBOMOLE} (their no-resolution of identity, noRI result)
in both cases using a def2-QZVP Gaussian basis set, which is less converged with respect to
our \textit{d-aug}-cc-pV5Z\@.
By using the same def2-QZVP basis set we were able to reproduce their same result:
23.4769 eV\@.
Van Setten \textit{et al.}\ also quote a plane waves $G_0W_0$ result by the \textsc{Berkeley-$GW$} code
using a plasmon-pole model and again starting from PBE: 24.10 eV, that is 0.6 eV larger than
the Gaussian basis result.
This result is very close to the Morris \textit{et al.}\ \cite{MorrisGodby07} result of 24.20 eV
obtained by a $G_0W_0$ on top of DFT LDA, using plane waves and with a full treatment
of the frequency dependence, i.e., without using the plasmon-pole model.
We remark that our $G_0W_0$ result starting from LDA is also larger (by 0.17 eV) than
the $G_0W_0$ starting from PBE\@.
So, our data seem coherent with the data available in the literature, in the limit
of the expected differences between using localized and delocalized basis sets.

\begin{figure}
 \includegraphics[width=\columnwidth]{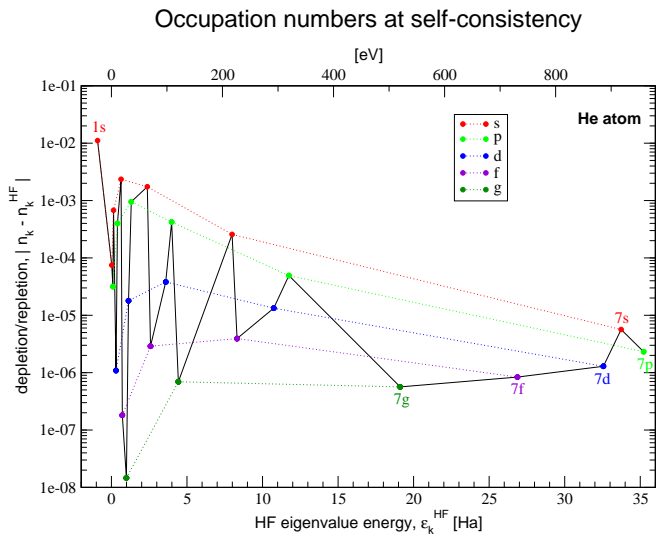}
 \caption{ Depletion and repletion of occupation numbers as calculated
   in r-RPA (renormalized RPA) towards SCRPA taken at
   self-consistency, as a function of the Hartree-Fock energies.  The
   orbital character of the states is indicated by different colors
   and as label of dots.  The zero of energy is set to the first
   ionization onset.  }
 \label{occupations}
\end{figure}

\subsection{Renormalized RPA (r-RPA) and single quasiparticle energies}

We will now present the self-consistent results  of our r-RPA
calculation.  Three or four iterations were necessary to achieve
self-consistency at the accuracy we quote in our tables.  In
Fig.\ \ref{occupations} we show the values of depletion/repletion of
the correlated r-RPA occupation numbers, $|n_k - n_k^\mathrm{HF}|$,
with respect to the integer uncorrelated HF occupation numbers, as a
function of the HF single-particle energy.  Our data were calculated
by Eq.\ (\ref{occ1}) and Eq.\ (\ref{occ2}) and include both $S=0$ spin
singlet and $S=1$ triplet contributions.  We remark that the
correlation corrections to the occupation numbers are small, 1\% for
the only occupied $1s$ level, becoming smaller and smaller for the
unoccupied ones with increasing principal quantum number $n$.  In the plot we
are also able to reveal a decreasing trend at increasing angular
momentum $l$.  In the jellium metallic spheres studied by Catara
\textit{et al.}\ \cite{CataraVanGiai96} depletions and repletions were
found much larger, beyond 30\% in some cases.  As already indicated in
the literature
\cite{OlevanoHolzmann12,HolzmannDelleSite11,HuotariOlevano10}, the
absolute value of depletions and repletions in occupation numbers and
momentum distributions, can be considered a reliable indication of the
correlation strength in one system.

\begin{table}[b]
 \begin{tabular}{lcccc}
   \hline \hline
   $nl$ & HF & $GW$ & Exact & r-RPA \\
   \hline
   $1s$ ($= - \mathrm{IP}$) & $-0.9179$ & $-0.9075$ & $-0.9037$ & $-0.9123$ \\
   $2s$ ($= - \mathrm{EA}$) & $+0.0217$ & $+0.0213$ & $> 0$ & $+0.0202$ \\
   $2p$ & $+0.0956$ & $+0.0944$ & & $+0.0935$ \\
   $3s$ & $+0.1394$ & $+0.1369$ & & $+0.1370$ \\
   \hline \hline
 \end{tabular}
 \caption{He electron removal (first line) and addition (following
   lines) energies (Ha) in HF, $GW$, exact \cite{KonoHattori84} and
   experimental (EXP) result and renormalized RPA (r-RPA)\@.
   The zero of energy is set to the first
   ionization onset, so that the ground-state value is coincident with
   minus the ionization potential.}
 \label{qpenergies}
\end{table}

In Table \ref{qpenergies} we report the calculated r-RPA
single-particle energies, as calculated by solving the single-particle
Schr\"odinger equation (\ref{MFeq}), using the mean-field potential
Eq.\ (\ref{MFpot}) calculated with the fractional correlated
occupation numbers already plotted in Fig.\ \ref{occupations}.  We report
the values at self-consistency and
compare them to the values calculated with other approaches as HF, $GW$,
and the exact values only where known, 
in practice just only the ionization potential can be derived
from an exact Hylleraas calculation. Focussing on
the last occupied $1s$ energy, we see that the 1.6\% error of HF is
reduced to less than 1\% in r-RPA, showing the same correct trend as the
$GW$ correction which reduces the error to 0.4\%.  The Hylleraas
calculation cannot provide the exact values of the electron affinity
and other addition energies, but comparing the r-RPA values to HF and
$GW$ we see that, with respect to HF, they go in the same direction of
$GW$ corrections, and go even beyond them.  They are anyway very close
to $GW$ quasiparticle energies.  So, the correlation corrections
brought by both r-RPA and $GW$ on top of the HF electronic structure
seem to go in the same direction, although it is, \textit{a priori}, not clear how they
are physically related to each other. We may clarify this point in a future publication.
We remark in particular that all the HOMO-LUMO+n gaps
close down from HF, and r-RPA situates half way with respect to $GW$.

\begin{figure*}
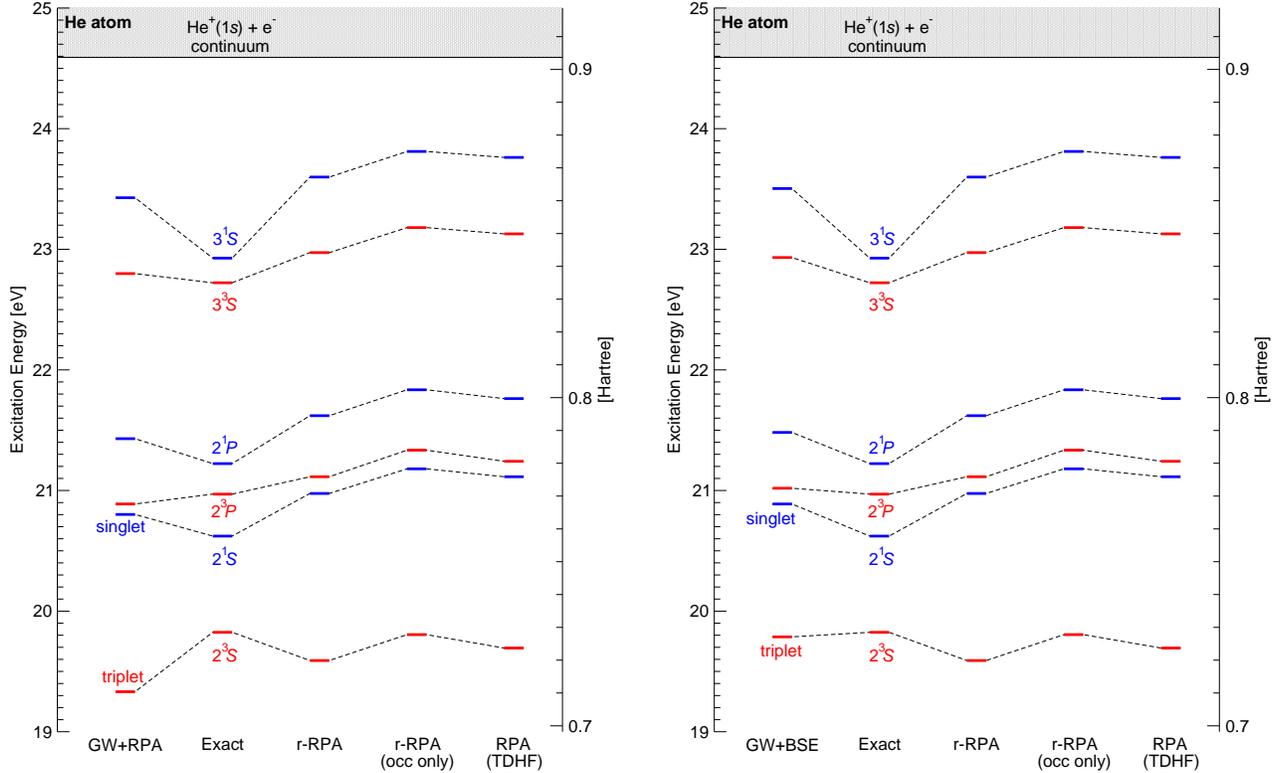

 \includegraphics[width=0.45\textwidth]{hegwrpaexactrrparpa}
 \hfil
 \includegraphics[width=0.45\textwidth]{hebseexactrrparpa}
 \caption{ He atom excitation spectrum by the renormalized r-RPA
   towards SCRPA, in two different flavors: updating up to
   self-consistency the occupation numbers only (r-RPA occ.\ only); and
   updating both occupation numbers and single-particle energies
   (r-RPA \textit{tout court}).  We compare the r-RPA spectra to the
   standard RPA, to the exact solution and finally also to a $GW$+RPA
   unscreened kernel approximation (left panel) and to a $GW$+BSE
   screened kernel calculation (right).  The zero of energy is set to
   the ground state $1^1\!S$.}
 \label{rrpa}
\end{figure*}

\begin{table}
\begin{tabular}{ccccccc}
 \hline \hline
 $n^S\!L$ & \parbox[c][3.5em][c]{.07\textwidth}{RPA\\(TDHF)} & \parbox[c]{.07\textwidth}{r-RPA\\occ.\\only} & \parbox[c]{.06\textwidth}{r-RPA\\occ.\ \&\\ene.}  & \parbox[c]{.06\textwidth}{\textbf{Exact}} &  \parbox[c]{.07\textwidth}{$GW$\\+\\RPA} &  \parbox[c]{.06\textwidth}{$GW$\\+\\BSE} \\
 \hline
 \multicolumn{6}{c}{atomic units [Ha]} \\
 $2^3\!S$ & 0.7237 & 0.7278 & 0.7199 & \textbf{0.7285} & 0.7104 & 0.7271 \\
 $2^1\!S$ & 0.7759 & 0.7783 & 0.7708 & \textbf{0.7578} & 0.7644 & 0.7676 \\
 $2^3\!P$ & 0.7806 & 0.7840 & 0.7759 & \textbf{0.7706} & 0.7676 & 0.7724 \\
 $2^1\!P$ & 0.7997 & 0.8024 & 0.7945 & \textbf{0.7799} & 0.7875 & 0.7894 \\
 $3^3\!S$ & 0.8499 & 0.8518 & 0.8442 & \textbf{0.8350} & 0.8378 & 0.8427 \\
 $3^1\!S$ & 0.8732 & 0.8750 & 0.8672 & \textbf{0.8425} & 0.8609 & 0.8637 \\
 \hline
 \multicolumn{6}{c}{electronvolt [eV]} \\
 $2^3\!S$ & 19.692 & 19.805 & 19.590 & \textbf{19.824} & 19.330 & 19.786 \\
 $2^1\!S$ & 21.115 & 21.178 & 20.976 & \textbf{20.621} & 20.800 & 20.888 \\
 $2^3\!P$ & 21.242 & 21.333 & 21.112 & \textbf{20.969} & 20.888 & 21.018 \\
 $2^1\!P$ & 21.762 & 21.835 & 21.619 & \textbf{21.222} & 21.428 & 21.480 \\
 $3^3\!S$ & 23.128 & 23.179 & 22.973 & \textbf{22.722} & 22.798 & 22.930 \\
 $3^1\!S$ & 23.762 & 23.811 & 23.598 & \textbf{22.926} & 23.426 & 23.502 \\
 \hline \hline
\end{tabular}
\caption{ He excitation energies in atomic units (Hartree, top) and
  electronvolt (eV, bottom) as calculated in a renormalized RPA
  towards SCRPA, both updating only the occupation numbers, or also
  the energies.  The zero of energy is set to the He ground state
  $1^1\!S$.  }
\label{excitationsrrpa}
\end{table}

In Fig.\ \ref{rrpa} and Table \ref{excitationsrrpa} we report on the
excitation energies obtained at self-consistency by the r-RPA
approximation.  We distinguish the case of updating only the
occupation numbers [Eqs.\ (\ref{occ1}) and (\ref{occ2})] keeping the
energies at the level of HF (indicated in the table and in the figure
as ``r-RPA occ.\ only''), from the full r-RPA, where we update
occupation numbers and energies [Eq.\ (\ref{MFeq}), indicated in figures
and tables as ``r-RPA occ.\ \& ene.''].  In all the cases we report the
result at self-consistency.  By looking at Fig.\ \ref{rrpa} and
Table \ref{excitationsrrpa} we see that the introduction of correlated
occupation numbers (r-RPA occ.\ only) systematically increases the
energy of all excitations with respect to standard RPA\@.  This results
in an overall worsening of the excitation spectrum with respect to the
exact.  Renormalized RPA (r-RPA) improves only on the first $2^3\!S$ excitation
whose energy is the only one underestimated by standard RPA\@.  This
result is rather discouraging since, when looking at the full SCRPA
matrix $\mathcal{S}$, it can be expected that the introduction of
fractional occupation numbers should play a major role in SCRPA\@.

However the situation is completely reversed when considering a full
r-RPA, taking into account corrections to the occupation numbers and
also to energies (r-RPA occ.\ \& ene.\ or r-RPA \textit{tout court}).
The effect of replacing occupation numbers in energies, that might
appear second-order with respect to their direct effect when replaced
where they appear in the $\mathcal{S}$ matrix, is instead quite
important to correct standard RPA towards the right direction.  We see
that the excitation energy is reduced not only with respect to the
r-RPA occ.\ only approximation, but also with respect to standard RPA\@.
This results in an overall improvement with respect to standard RPA,
towards the exact solution.  Again the exception is represented by the
first excited state where in full r-RPA we observe a worsening.

These r-RPA results can be better understood if compared not directly
with the $GW$+BSE approach, but rather with a $GW$+RPA calculation using a
$\bar{v}$ unscreened kernel.  Indeed, in both the full r-RPA and the
$GW$+RPA cases the novelty with respect to standard RPA (TDHF) is the
introduction of a correlated, in place of the uncorrelated HF
electronic structure, as starting point of the RPA equations.  While
the kernel keeps in all cases the same $\bar{v}$ as in standard RPA\@.
We see in Fig.\ \ref{rrpa} (left panel) and Table \ref{excitationsrrpa}
that, with respect to standard RPA, the effect of both $GW$+RPA and
r-RPA is exactly in the same direction.  For all excitations we
observe a reduction of their energy with respect to standard RPA\@.
This can be directly traced back to the reduction of single-particle
HOMO-LUMO gaps taken as starting points to the same $\bar{v}$ kernel
RPA equations.  The $GW$+RPA excitation energies are lower than r-RPA
simply because the HOMO-LUMO $GW$ gaps are smaller than r-RPA\@.  For
this reason the $GW$+RPA is more in agreement with the exact result,
again with the exception of the first $2^3\!S$ excitation where both
r-RPA and $GW$+RPA go in the wrong direction with respect to standard
RPA, and the more important $GW$+RPA correction turns out in a worse
result.

To correctly describe this first excitation and improve, rather than
worsen with respect to standard RPA, a correction of the kernel seems
required.  For example the introduction of screening into the bare
particle-hole exchange interaction of the RPA kernel, like done in BSE\@.
This reduces the strength of the kernel and so of the correction to
the excitation energy when starting from $GW$+dRPA\@.  The effect of the
screened BSE $W$ kernel is impressively evident on this first $2^3\!S$
excitation (compare left and right panel of Fig.\ \ref{rrpa}).  The
screening reduces its effect when moving to higher excitations.  For
the highest excitations one might argue that the introduction of the
screening, although with smaller and smaller effect, goes in the wrong
direction to increase the energy, but we remind that the
overestimation of the excitation energy is a finite basis set effect
due to the poor representation of highly delocalized states by
Gaussians also detected in the CI calculation.

The r-RPA result here presented may appear  not yet
satisfactory, for example if compared to BSE\@. However, we think it is a very
encouraging result.   This result  makes us 
hope that the introduction of the two-particle correlation terms into
the full SCRPA $\mathcal{S}$ matrix can reduce the strength of the
kernel, like it happens in BSE when introducing the screening into the
bare Coulomb $v$. 
Indeed, the neglect of the correlation terms in $\mathcal{S}$ atrophies SCRPA very much. 
This the more so as the correlation terms can be shown to contain screening in a similar way as with BSE\@. 
These aspects may be elaborated in a future publication.

In Table \ref{rrpaos} we report the $f_{1^1\!S \to 2^1\!P}$ first
dipole allowed transition oscillator strength for r-RPA\@.  We remark an
improvement with respect to standard RPA\@.  This is mostly due to the
update of occupation numbers.  Since the oscillator strength is above
all sensitive to wave functions, the difference between r-RPA with or
without updating the energies is less evident than in excitation
energies themselves.  Nevertheless, the fact to have different
energies along the diagonal of the RPA $\mathcal{S}$ matrix has also
an effect on eigenvectors, wave functions and, thus,  oscillator strengths.
This effect is also appreciable when comparing the standard RPA to the
$GW$+RPA oscillator strength.  The latter even shows a worsening.  A
correction to occupation numbers and/or the kernel, like in BSE, is
required to improve the oscillator strength towards the good direction.

\begin{table}[t]
\begin{tabular}{ccccccc}
  \hline \hline
   & \parbox[c][3.5em][c]{.06\textwidth}{RPA\\(TDHF)} & \parbox[c]{.06\textwidth}{r-RPA\\occ.\\only} & \parbox[c]{.06\textwidth}{r-RPA\\occ.\ \&\\ene.}  & \parbox[c]{.06\textwidth}{\textbf{Exact}} &  \parbox[c]{.05\textwidth}{$GW$\\+\\RPA} &  \parbox[c]{.05\textwidth}{$GW$\\+\\BSE} \\
  \hline
  $f_{1^1\!S \to 2^1\!P}$ & 0.2916 & 0.2889 & 0.2877 & \textbf{0.27616} & 0.2946 & 0.2763 \\ 
  \hline \hline
\end{tabular}
\caption{Helium atom first dipole-allowed $1^1\!S \to 2^1\!P$ transition 
  oscillator strength, calculated in RPA (TDHF), r-RPA updating up to
  self-consistence occupation numbers only, r-RPA updating both
  occupation numbers and energies, exact Hylleraas calculation
  \cite{KonoHattori84}, RPA on top of $GW$ quasiparticle energies, and
  BSE\@.  }
 \label{rrpaos}
\end{table}

\section{Conclusions}

Our work presented a comparison on the same footing,
in particular using the same Gaussian basis set,
of several many-body approaches,
including a not so much explored renormalized RPA (r-RPA) derived
from the EOM method developed in nuclear physics.
Our work shows that the r-RPA, which is a sub-product of the
SCRPA approach, improves over the standard RPA (i.e.\ linearized
time-dependent Hartree-Fock (TDHF) \cite{Rowe66}) and achieves a
result of accuracy comparable to $GW$+BSE, except for the first
excited state where there is no improvement.  Also $GW$+BSE improves
on dRPA on top of both HF and $GW$, but also on RPA/TDHF\@.  CI is
certainly one of the most accurate methods, but localized-basis-set
issues seriously reduce its accuracy on the highest excited states,
well outside chemical accuracy.  On the Rydberg series, an
Exact-DFT+TDLDA calculation done in real space shows superior
performances with respect to even Gaussian-basis CI\@.  
In the same CI Gaussian basis set, we have presented also the 
DFT-LDA+dRPA and DFT-LDA+TDLDA 
helium excitation spectra, arguing that the question of the
boundness of the Rydberg series depends on the way the ionization
potential is calculated.
On the ground state CI achieves chemical accuracy, but
cannot do better even relying on recent basis set extrapolation
techniques.  On the other hand, standard QMC, Slater-Jastrow
variational Monte Carlo (VMC) followed by diffusion Monte Carlo (DMC)
at the actual computer power, has shown 2 orders of magnitude superior
accuracy with respect to CI\@.  We should mention however that the
helium ground state wave function is nodeless, a favorable case where
QMC is unaffected by the so-called fermion sign problem.

\begin{acknowledgments}
We thank Xavier Blase, Ivan Duchemin, and Markus Holzmann for useful discussions. 
\end{acknowledgments}

\appendix

\bibliographystyle{SciPost_bibstyle}
\bibliography{he}

\end{document}